\documentclass[fleqn,usenatbib]{mnras}
\usepackage{newtxtext,newtxmath}
\usepackage[T1]{fontenc}
\DeclareRobustCommand{\VAN}[3]{#2}
\let\VANthebibliography\thebibliography
\def\thebibliography{\DeclareRobustCommand{\VAN}[3]{##3}\VANthebibliography}

\usepackage{graphicx}
\usepackage{amsmath}
\usepackage{subcaption}
\usepackage{multicol}
\usepackage{xspace}
\newcommand{\bemu}{{\tt Baccoemu}\xspace}

\title[Calibrating baryonic effects in cosmic shear]{Calibrating baryonic effects in cosmic shear with external data in the LSST era}

\author[A. Wayland et al.]{
Amy Wayland\thanks{E-mail: amy.wayland@physics.ox.ac.uk}, David Alonso, and Matteo Zennaro
\\
Department of Physics, University of Oxford, Denys Wilkinson Building, Keble Road, Oxford OX1 3RH, United Kingdom
}

\date{Accepted XXX. Received YYY; in original form ZZZ}
\pubyear{\the\year{}}

\begin{document}
\label{firstpage}
\pagerange{\pageref{firstpage}--\pageref{lastpage}}
\maketitle

\begin{abstract}
  Cosmological constraints derived from weak lensing (WL) surveys are limited by baryonic effects, which suppress the non-linear matter power spectrum on small scales. By combining WL measurements with data from external tracers of the gas around massive structures, it is possible to calibrate baryonic effects and, therefore, obtain more precise cosmological constraints. In this study, we generate mock data for a Stage-IV weak lensing survey such as the Legacy Survey of Space and Time (LSST), X-ray gas fractions, and stacked kinetic Sunyaev-Zel'dovich (kSZ) measurements, to jointly constrain cosmological and astrophysical parameters describing baryonic effects (using the Baryon Correction Model -- BCM). First, using WL data alone, we quantify the level to which the BCM parameters will need to be constrained to recover the cosmological constraints obtained under the assumption of perfect knowledge of baryonic feedback. We identify the most relevant baryonic parameters and determine that they must be calibrated to a precision of $\sim 10$-$20\%$ to avoid significant degradation of the fiducial WL constraints. We forecast that long-term X-ray data from $\mathcal{O}(5000)$ clusters should be able to reach this threshold for the parameters that characterise the abundance of hot virialised gas. Constraining the distribution of ejected gas presents a greater challenge, however, but we forecast that long-term kSZ data from a CMB-S4-like experiment should achieve the level of precision required for full self-calibration.
\end{abstract}



\section{Introduction}
  The distribution of baryonic matter on small scales is strongly affected by complex astrophysical processes, which are collectively referred to as baryonic feedback \citep{vanDaalen:2011,Chisari:2019}. These processes, including gas cooling, star formation, and feedback from active galactic nuclei (AGN), can substantially modify the matter distribution within dark matter haloes. On non-linear scales, baryonic feedback leads to a significant suppression of the matter power spectrum. For instance, AGN feedback redistributes baryons to the outer regions of haloes, leading to an overall lower matter density on halo-sized scales. Cosmological simulations indicate that these effects lead to suppression of the power spectrum by $O(10\%)$ on scales in the range $0.1 \lesssim k \lesssim 10 h \text{Mpc}^{-1}$ \citep{Steinborn:2015, Chisari:2019, Schaye:2023, Pakmor:2022}. In contrast, star formation in the centres of haloes results in a contraction of the matter profile, enhancing the power spectrum on the smallest scales ($k \gtrsim 10 h \text{Mpc}^{-1}$). Baryonic effects lead to a non-trivial modification of the underlying matter distribution that is difficult to model precisely without direct observations \citep{Crain:2023}. As a result, the impact of baryonic feedback on the matter power spectrum introduces systematic uncertainties that hinder the precise determination of cosmological parameters from weak lensing surveys \citep{vanDaalen:2011, Chisari:2019, Arico:2023,Garcia-Garcia:2024,Bigwood:2024}. 

  Weak gravitational lensing allows us to directly map the overdensity of matter fluctuations, providing a direct means to trace the history of the growth of structure in the Universe. However, the sensitivity of weak lensing to the matter distribution on small scales is limited by the uncertainties surrounding baryonic feedback, which in turn decreases the precision of cosmological constraints \citep{DES:2021,KiDS:2023}. To mitigate the systematic errors introduced by these effects, many current weak lensing analyses exclude the data from small scales altogether \citep[e.g.][]{Prat:2021, Zacharegkas:2021, Amon:2022, Lange:2023}. On the other hand, the Kilo-Degree Survey (KiDS) and Hyper Suprime-Cam (HSC) mitigate uncertainties due to feedback by modelling the small scales with \texttt{HMcode} \citep{Mead:2020, Dalal:2023, KiDS:2023}. Incorporating these non-linear scales is essential to fully benefit from the cosmological potential of upcoming Stage-IV weak lensing surveys, such as the Legacy Survey of Space and Time (LSST) at the Vera C. Rubin Observatory \citep{LSST:2008,2018arXiv180901669T}, or the Euclid satellite \citep{Euclid:2011}.

  Furthermore, the $S_8$ tension, which is the discrepancy between the constraints on the clustering amplitude parameter $S_8 = \sigma_8 \sqrt{\Omega_{\mathrm{m}}/0.3}$ inferred from some weak lensing surveys and those obtained from the CMB, could be explained by a suppression of the non-linear matter power spectrum \citep{Schneider:2021,2206.11794,DES:2022,Arico:2023,McCarthy:2023}. This suppression may be indicative of either stronger baryonic feedback than predicted by hydrodynamical simulations, or a departure from the standard cosmological model. Recent results from the KiDS collaboration point instead to other causes related to the calibration of redshift distributions and shape measurements \citep{Wright:2025}. Independent of the cause, it is essential to improve our understanding of baryonic feedback in order to determine the origin of the $S_8$ tension and to ensure the robustness of any future cosmological constraints from weak lensing.

  External tracers of the large-scale structure, such as X-ray gas fractions and measurements of the kinematic Sunyaev-Zel'dovich (kSZ) effect, can offer valuable complementary information to help us constrain baryonic feedback. X-ray gas fractions, which measure the fraction of baryons in the bound gas component of galaxy clusters as a function of halo mass, are a direct probe of the hot baryon content within haloes. Meanwhile, kSZ measurements, which capture the motion and abundance of ionised gas through the scattering of hot electrons off CMB photons, allow us to trace both the bound and ejected gas components of haloes. These external tracers can help refine our models of baryonic feedback, providing additional constraints on both the cosmological and baryonic parameters \citep{Schneider:2018,Schneider:2021,Grandis:2023,Bigwood:2024,To:2024,McCarthy:2024,Sunseri:2025}.

  At present, it remains uncertain whether external data will be sufficient to constrain baryonic feedback parameters to the level required by the next-generation of cosmic shear surveys \citep{2308.01856,2411.04088}. In this work, we use mock LSST-like weak lensing data to determine the level to which each baryonic parameter needs to be calibrated to recover optimal cosmological constraints. We then investigate whether near-term and long-term X-ray and kSZ data have the potential to achieve these calibration requirements. Finally, we combine all three datasets (WL + X-rays + kSZ) to quantify the improvement in cosmological constraints over WL alone.

  The structure of the paper is as follows. In Section \ref{sec:model}, we describe the theoretical model for cosmic shear, baryonic effects, X-ray gas fractions, and the stacked kSZ profile. The results of our analysis are presented in Section \ref{sec:results}, where we forecast the constraints required on baryonic parameters, before subsequently examining whether weak lensing combined with external tracers can achieve these constraints. Finally, we conclude with a summary of our key findings and their implications for future cosmological surveys in Section \ref{sec:conclusion}.


\section{Cosmic shear and baryonic effects} \label{sec:model}
  \subsection{The cosmic shear signal} \label{ssec:model.cshear}
    Weak gravitational lensing is the deflection of photon trajectories as they propagate along the line-of-sight from a background source to the observer, caused by the gravitational potential of the intervening large-scale structure \citep{Bartelmann:1999,Mandelbaum:2017}. This effect in turn leads to coherent distortions in the observed shapes of background galaxies, which is referred to as cosmic shear. As an unbiased tracer of the matter overdensity field, cosmic shear provides a direct probe of the large-scale distribution of matter in the Universe. 

    We model the cosmic shear signal using the Limber approximation \citep{Limber:1954}, under which the cross-correlation of the $E$-mode cosmic shear signal between two redshift bins, $i$ and $j$, in harmonic space, is related to the 3D matter power spectrum, $P_{\mathrm{mm}}(k,z)$, as
    \begin{equation} \label{eq:cosmic_shear_C_ell}
      C_{\ell}^{ij} = G_\ell^2 \int \frac{\mathrm{d} \chi}{\chi^2} \, q_i(\chi) q_j(\chi) P_{\mathrm{mm}}\left(k=\frac{\ell + 1/2}{\chi}, \, z(\chi) \right),
    \end{equation}
    where $\ell$ is the angular multipole, $k$ is the wavenumber, $z$ is the redshift, $\chi$ is the comoving radial distance, and the prefactor $G_\ell$ accounts for the difference between the 3D Laplacian of the gravitational potential and the angular Hessian of the associated lensing potential \citep{Kilbinger:2017},
    \begin{equation}
      G_\ell \equiv \sqrt{\frac{(\ell+2)!}{(\ell-2)!}} \frac{1}{(\ell+1/2)^2}.
    \end{equation}
    The lensing kernel $q_i$ of the $i$-th redshift bin is related to the redshift distribution of the sample, $p_i(z)$, via
    \begin{equation} \label{eq:radial_kernel}
      q_i(\chi) = \frac{3}{2} H_0^2 \Omega_{\mathrm{m}} \frac{\chi}{a(\chi)} \int_{z(\chi)}^\infty \mathrm{d}z' \, p_i(z') \frac{\chi(z') - \chi}{\chi(z')},
    \end{equation}
    where $a = 1/(1+z)$ is the scale factor. 

    In this work, all theory calculations are carried out using the Core Cosmology Library \citep[CCL,][]{CCL}. To accelerate the evaluation of the theory predictions, we use the \texttt{Halofit} parametrisation to model the dark matter-only non-linear power spectrum \citep{Smith:2002, Takahashi:2012}. We include the suppression due to baryonic effects using the baryonification approach as implemented in \bemu \citep{Angulo:2020, baccoemu}, with the specific parametrisations given in Section \ref{ssec:model.bcm}. We note that we ignore the redshift evolution of the baryonic suppression, and generate synthetic data with one set of baryonic parameters applied to the redshift dependent non-linear power spectra. While the baryonic parameters should in principle evolve across redshift bins, analyses such as \citet{Garcia-Garcia:2024} have shown that current observational datasets lack the sensitivity required to robustly constrain such evolution. For simplicity, and without hindering the generality of our analysis, we adopt the same assumption for our LSST-like synthetic data.

    To generate our mock cosmic shear data, we follow the description of the LSST Dark Energy Science Collaboration Science Requirements Document \citep{1809.01669} for the 10-year weak lensing sample. We assume a number density of sources, redshift distribution, and photometric redshift uncertainty as described in \citet{2212.04291}, and we divide the sample into the same five equal-density redshift bins. The cosmic shear data vector then consists of all the auto- and cross-power spectra between the all redshift bin pairs.

  \subsection{Systematics affecting the cosmic shear signal} \label{ssec:model.systematics}
    The cosmic shear signal is affected by various sources of systematic uncertainty, which can be categorised into two classes: calibratable and non-calibratable parameters. Tight priors can be imposed on the calibratable systematics through independent external observations or by calibrating the instrument measurements. In contrast, non-calibratable systematics can only be constrained by the data itself. 

    \subsubsection{Intrinsic alignments}
      The intrinsic alignment (IA) of galaxies with their local environment is a non-calibratable systematic effect that impacts cosmic shear data \citep{Brown:2002}. The simplest physical model for intrinsic alignments is the tidal alignment model, more commonly called the non-linear linear alignment model \citep[NLA][]{Hirata:2004}. Consider a nearby structure at the same redshift as a sample of galaxies. The gravitational tidal field is stronger at the end of a galaxy closest to the structure and weaker at the opposite end. This results in a distortion in the shape of the galaxy, causing it to elongate and align with the structure along the longitudinal direction. The NLA model leads to a contribution to the shear power spectrum that is proportional to the matter power spectrum integrated over the redshift distributions of the samples being correlated. The strength of the effect is quantified through a redshift-dependent amplitude, commonly parametrised as
      \begin{equation} \label{eq:NLA_IA}
        A_{\mathrm{IA}}(z) = A_{\mathrm{IA},0} \left(\frac{1+z}{1+z_*}\right)^{\eta_{\mathrm{IA}}}.
      \end{equation}
      Here $A_{\mathrm{IA},0}$ and $\eta_{\mathrm{IA}}$ are free parameters that characterise the amplitude and slope of the redshift power-law, respectively. We use $z_* = 0.62$ as in \citet{DES:2017troxel} and \citet{DES:2017abbott}. Hence, using the NLA model to account for intrinsic alignments introduces two additional free parameters to marginalise over into the likelihood function.

      More complex physical process may also contribute to the alignment of galaxies with the large-scale structure, and more general models exist to describe them. These include, for instance, the Tidal Alignment and Tidal Torquing model \citep[TATT][]{Blazek2019}, as well as halo models targeting IAs on small scales \citep{Fortuna2021}. Given the relatively low amplitude of IAs found in current weak lensing samples \cite{DES:2021secco,Wright:2025}, and the reduced accuracy with which the effect needs to be modelled \citep{Paopiamsap:2023}, we limit our analysis to the NLA parametrisation for simplicity. Nevertheless, we acknowledge that the NLA model represents one of the simplest alternatives in practice. Other methods have been proposed in the literature to assess the robustness of cosmological constraints derived from weak lensing. One such approach is blue-only analysis, which minimises the impact of intrinsic alignments \citep{McCullough:2024, Siegel:2025}.

    \subsubsection{Photometric redshifts and multiplicative bias} \label{sssec:model.photo_z_and_multbias}
      Photometric redshift uncertainties and multiplicative shape measurement biases are important calibratable systematics that affect cosmic shear data \cite{Bonnett:2016,Hildebrandt:2018}. The true redshift distributions of the galaxy samples are subject to uncertainty due to the lack of precise redshift measurements in photometric surveys, with the associated errors referred to as photometric redshift uncertainties. The uncertainty in the distribution of the $i$-th redshift bin can be characterised by a parameter $\Delta z_i$ that shifts the mean of the redshift bin. This has been shown to capture the impact of redshift uncertainties on cosmic shear data with sufficient accuracy \citep{Bonnett:2016,Ruiz-Zapatero:2023}. The true redshift distribution is then given by $p_i(z) = \hat{p}_i(z + \Delta z_i)$, where $\hat{p}_i(z)$ represents the best-guess redshift distribution \citep{Ruiz-Zapatero:2023}.

      In addition to photometric redshift uncertainties, cosmic shear data is also affected by biases in the measured shape of galaxies arising from the limited resolution and image noise \citep{Miller:2013}. This bias is commonly characterised in terms of a multiplicative factor $m_i$ for the $i$-th sample, known as the multiplicative bias \citep[see e.g.][]{2017MNRAS.465.1454H}. Specifically, the measured angular power spectra, $\widetilde{C}_\ell$, are related to the true angular power spectra, $C_\ell$, via
      \begin{equation} \label{eq:multiplicative_bias}
        \widetilde{C}_\ell^{ij} = (1+m_i)(1+m_j) C_\ell^{ij},
      \end{equation}
      where $m_i$ and $m_j$ are the multiplicative biases of the redshift bins $i$ and $j$, respectively.

      We can marginalise over the calibratable parameters using an analytical method based on the Laplace approximation, as proposed in \citet{Ruiz-Zapatero:2023,Hadzhiyska:2023}. This approach linearises the dependence of the theoretical prediction for the data with respect to the calibratable parameters. As a result, the covariance matrix is modified such that the linear combinations of the data most sensitive to variations in these parameters are assigned higher variance. The key advantage of this method is that it avoids introducing two additional free parameters, $\Delta z_i$ and $m_i$, per redshift bin $i$ into the likelihood function, thus significantly improving the computational efficiency of running MCMC chains. Furthermore, \citet{Ruiz-Zapatero:2023} demonstrated that analytical marginalisation recovers the same constraints on the cosmological parameters as directly sampling the full parameter space, even when marginalising over uncertainties in the full shape of the redshift distribution, instead of shifts in its mean.

      To implement the analytical marginalisation numerically, we simply update the covariance matrix to
      \begin{equation} \label{eq:updated_cov}
        \widetilde{\mathsf{C}} \equiv \mathsf{C} + \mathsf{T}\,\mathsf{P}\,\mathsf{T}^T,
      \end{equation}
      where $\mathsf{P}$ is the covariance matrix of the calibratable systematics, which we assume to be uncorrelated, and the matrix $\mathsf{T}$ contains the derivatives of the theoretical prediction $\mathbf{t}$ with respect to the set of calibratable parameters $\mathbf{v}$, evaluated at their prior mean values, $\overline{\mathbf{v}}$,
      \begin{equation}
        \mathsf{T} \equiv \left. \frac{\mathrm{d} \mathbf{t}}{\mathrm{d} \mathbf{v}} \right|_{\mathbf{v} = \bar{\mathbf{v}}}.
      \end{equation}
      In our analysis, we assume a Gaussian prior on both $m_i$ and $\Delta z_i$, with 68$\%$ uncertainties $\sigma(m_i) = 0.01$ and $\sigma(\Delta z_i) = 0.001 (1+\overline{z}_i)$, where $\overline{z}_i$ is the mean of the $i$th redshift bin. These calibration priors correspond to the LSST requirements as reported in \citet{2018arXiv180901669T}.

  \subsection{The baryon correction model}\label{ssec:model.bcm}
    The other main source of non-calibratable systematic uncertainty for cosmic shear is baryonic feedback. To characterise this effect, we use the baryon correction model (BCM), developed in \citet{Schneider:2015}. In this model, each halo is decomposed into four components: adiabatically relaxed dark matter, bound gas in hydrodynamical equilibrium, gas ejected from feedback processes, and stars in the central galaxy. This framework allows for a direct connection between the total matter density and the observed distribution of gas and stars in haloes, in turn enabling the physics-based study of baryonic effects on the matter power spectrum. 

    The BCM modifies the dark matter density profile of haloes to account for the presence of gas and stars via:
    \begin{equation}
      \rho_{\mathrm{dmo}}(r) \rightarrow \rho_{\mathrm{dmb}}(r) = \rho_{\mathrm{cdm}}(r) + \rho_{\mathrm{gas}}(r) + \rho_{*}(r).
    \end{equation}
    Here, $\rho_{\mathrm{dmo}}$ represents the dark matter density profile in the absence of baryonic effects, while $\rho_{\mathrm{dmb}}$ denotes the dark-matter-baryon profile, which is in turn composed of three components as follows. The cold dark matter profile, $\rho_{\mathrm{cdm}}$, is calculated by modifying the standard Navarro-Frenk-White (NFW) profile \citep{Navarro:1997} to allow for the adiabatic expansion and relaxation of cold dark matter. The total gas profile, $\rho_{\mathrm{gas}}(r)$, is composed of both bound and ejected gas components, as described in Sections \ref{sssec:model.bgas} and \ref{sssec:model.egas}. Baryonic effects, such as AGN feedback, play a significant role in altering the gas distribution around haloes over cosmological scales. The stellar profile, $\rho_{*}(r)$, is characterised by a power law with an exponential cut-off, which primarily influences the central region of the halo, rather than larger cosmological scales. Within the framework of the BCM, particles in each halo from a dark-matter only simulation are displaced according to the relative differences between the theoretical baryonic and non-baryonic profiles. As a consequence, the original matter distribution is corrected to account for baryonic effects, while preserving the triaxial structure induced by gravity in the simulation.

    In this work, we use the emulator \bemu introduced in \citet{baccoemu}, which implements the BCM over a wide parameter space, to simultaneously constrain cosmology and baryonic feedback. The model contains seven free baryonic parameters that are connected to the gas and stellar profiles within dark matter haloes. This framework allows us to establish the relationship between the degree of suppression of the power spectrum due to baryonic feedback and the gas and stellar fractions in haloes. In the following sections, we present the mass fractions and density profile models that are subsequently used to calculate the bound gas fraction inferred from X-ray data and the temperature shift of the CMB resulting from the kSZ effect. Specifically, we use the dark matter profile and ejected gas profile of \citet{Schneider:2015}, the bound gas profile of \citet{Arico:2020a}, and the stellar profile of \citet{Behroozi:2013}.

    \subsubsection{Dark matter}
      The dark matter profile is assumed to follow the NFW profile of \citet{Navarro:1997}. Following \citet{Baltz:2009} and \citet{Oguri:2011}, we truncate the NFW profile at radius $r_{\mathrm{tr}}$ to prevent the total mass diverging,
      \begin{equation} \label{eq:rho_nfw}
        \rho_{\mathrm{nfw}}(x,y) = \frac{\rho_0}{x(1+x)^2} \frac{1}{(1+y^2)^2},
      \end{equation}
      where $x \equiv r/r_{\mathrm{s}}$ and $y \equiv r/r_{\mathrm{tr}}$. Here, $r_{\mathrm{s}}$ denotes the scale radius, which is related to the halo virial radius $r_{200\mathrm{c}}$ through $r_{\mathrm{s}} \equiv  r_{200\mathrm{c}}/ c$, where $c$ is the concentration parameter. The truncation radius, $r_{\mathrm{tr}}$, marks the outer boundary of the halo. \citet{Schneider:2015} evaluated the truncated NFW profile using haloes from DM-only simulations and found that the best-fit value for the dimensionless truncation parameter $\tau \equiv r_{\mathrm{tr}} / r_{\mathrm{s}}$ is $\tau = 8c$. Consequently, we adopt $r_{\mathrm{tr}} = 8 r_{200\mathrm{c}}$ in our analysis.

    \subsubsection{Central galaxy}
      We adopt the parametrisation from \citet{Behroozi:2013} for the mass fraction of stars in the central galaxy, which is based on abundance matching,
      \begin{equation} \label{eq:f_*}
        f_*(M_{200\mathrm{c}}) = \epsilon \left(\frac{\mathcal{M}_1}{M_{200\mathrm{c}}}\right) 10^{g(\log_{10}(M_{200\mathrm{c}}/\mathcal{M}_1)) - g(0)},
      \end{equation}
      where the function $g(x)$ is defined as
      \begin{equation}
        g(x) = -\log_{10}(10^{\alpha x}+1) + \frac{\delta (\log_{10}(1+e^x))^\gamma}{1+e^{10^{-x}}}.
      \end{equation}
      In this expression, $\alpha$, $\gamma$, $\delta$, and $\epsilon$ are redshift-dependent parameters, with their functional form given in \citet{Behroozi:2013}. The free parameter $\mathcal{M}_1$ is related to the \bemu parameter $M_1$ via
      \begin{equation}
        \log_{10}(\mathcal{M}_1) = \log_{10} M_1 + ((a-1) \log_{10} M_{1,a} + z \log_{10} M_{1,z})\nu,
      \end{equation}
      where we use the values of the parameters $M_{1,a}$ and $M_{1,z}$ given in \citet{Behroozi:2013} and $\nu = \exp(-4a^2)$. Hence, the stellar component depends on a single free parameter, $\log_{10} M_1$, which represents the characteristic mass of the central galaxy of the halo at redshift $z=0$.

    \subsubsection{Bound gas} \label{sssec:model.bgas}
      The mass fraction of the bound gas as a function of the halo mass is parametrised as
      \begin{equation} \label{eq:f_bgas}
        f_{\mathrm{bgas}}(M_{200\mathrm{c}}) = \frac{\Omega_{\mathrm{b}}/\Omega_{\mathrm{m}} - f_*}{(1+(M_{\mathrm{c}}/M_{200\mathrm{c}})^\beta)},
      \end{equation}
      where $M_{\mathrm{c}}$ and $\beta$ are free parameters. Notably, $f_{\mathrm{bgas}} = \frac{1}{2}(\Omega_{\mathrm{b}}/\Omega_{\mathrm{m}} - f_*)$ at $M_{\mathrm{c}} = M_{200\mathrm{c}}$, which means that $M_{\mathrm{c}}$ represents the halo mass at which half of the gas has been ejected from the halo. The parameter $\beta$ describes the rate at which the gas is depleted towards smaller halo masses.

      We adopt a flexible form of the bound gas density profile, as introduced by \citet{Arico:2020a}, in which the profile is parametrised as a double power-law for $r<r_{\mathrm{out}}$, 
      \begin{equation} \label{eq:rho_bgas}
        \rho_{\mathrm{bgas}}(r < r_{\mathrm{out}}) = \frac{y_0}{(1+r/r_{\mathrm{inn}})^{\beta_{\mathrm{inn}}}} \frac{1}{(1+(r/r_{\mathrm{out}})^2)^2}.
      \end{equation}
      Here, $\beta_{\mathrm{inn}} = 3 - (M_{\mathrm{inn}}/M_{200\mathrm{c}})^{\mu_{\mathrm{inn}}}$ where $\mu_{\mathrm{inn}} = 0.31$ and $M_{\mathrm{inn}}$ is a free parameter that characterises the transition mass of the density profile of the hot gas in haloes. The characteristic scales $r_{\mathrm{inn}}$ and $r_{\mathrm{out}}$ describe how the slope of the profile evolves at small and large radii, respectively, and are defined as
      \begin{equation}
        r_{\mathrm{inn}} = \theta_{\mathrm{inn}} r_{200\mathrm{c}}  \text{ and } r_{\mathrm{out}} = \theta_{\mathrm{out}} r_{200\mathrm{c}},
      \end{equation}
      where $\theta_{\mathrm{inn}}$ and $\theta_{\mathrm{out}}$ are free parameters of the \bemu model. The normalisation constant $y_0$ in Equation \eqref{eq:rho_bgas} ensures that the integrated profile attains the correct mass $f_{\mathrm{bgas}} M_{200\mathrm{c}}$ at $r_{200\mathrm{c}}$.

      For radii beyond $r_{\mathrm{out}}$, we assume that the bound gas profile follows the truncated NFW profile given in Equation \eqref{eq:rho_nfw}. This assumption is physically motivated by the fact that the bound gas acts as a collisionless fluid in the outer regions of the halo. The normalisation constant $\rho_0$ in Equation \eqref{eq:rho_nfw} can be written as $\rho_0 = A y_0$ where the constant $A$ is determined by the requirement that the bound gas density profile is continuous at $r=r_{\mathrm{out}}$ and $y_0$ is the normalisation constant in Equation \eqref{eq:rho_bgas}.

      Hence, from Equations \eqref{eq:f_bgas} and \eqref{eq:rho_bgas}, we note that the bound gas component is governed by a total of five free parameters: $M_{\mathrm{c}}$, $\beta$, $\theta_{\mathrm{out}}$, $\theta_{\mathrm{inn}}$, and $M_{\mathrm{inn}}$.

    \subsubsection{Ejected gas} \label{sssec:model.egas}
      The ejected gas component represents the remaining gas that has neither been transformed into stars nor is part of the bound gas component, but has instead been mostly expelled from the halo by AGN-driven outflows. Hence, the mass fraction of ejected gas is given by
      \begin{equation} \label{eq:f_egas}
        f_{\mathrm{egas}}(M_{200\mathrm{c}}) = \Omega_{\mathrm{b}}/\Omega_{\mathrm{m}} - f_{\mathrm{bgas}}(M_{200\mathrm{c}}) - f_*(M_{200\mathrm{c}}),
      \end{equation}
      where $\Omega_{\mathrm{b}}/\Omega_{\mathrm{m}}$ is the total baryon mass fraction, $f_{\mathrm{bgas}}$ is the bound gas mass fraction given in Equation \eqref{eq:f_bgas}, and $f_*$ is the stellar mass fraction given in Equation \eqref{eq:f_*}. Assuming that the energy released by the AGN induces shifts in the velocities of the gas particles that follow a Maxwell-Boltzmann distribution, the ejected gas profile takes the form
      \begin{equation} \label{eq:rho_egas}
        \rho_{\mathrm{egas}}(r) = \frac{M_{200\mathrm{c}} f_{\mathrm{egas}}}{(2\pi r_{\mathrm{ej}}^2)^{3/2}} \exp\left[-\frac{1}{2} \left(\frac{r}{r_{\mathrm{ej}}}\right)^2\right],
      \end{equation}
      where the ejection radius $r_{\mathrm{ej}}$ is the maximum radius reached by the expelled gas, which we model as
      \begin{equation*}
        r_{\mathrm{ej}} \equiv \eta \times 0.75 \, r_{\mathrm{esc}}.
      \end{equation*}
      Here, $\eta$ is a free parameter and $r_{\mathrm{esc}}$ is the escape radius, which can be related to the virial radius of the halo, $r_{200\mathrm{c}}$ \citep{Schneider:2015}. Consequently, the ejected gas component depends only on a single free parameter, $\eta$.

      In summary, the Baryon Correction Model, as implemented in \bemu, incorporates a total of seven free baryonic parameters, as outlined in Table \ref{tab:free_parameters}. It is important to note, however, that the model does not account for potential complexities such as mass-dependent or redshift-dependent variations in these parameters. The impact of introducing redshift dependence is explored in Appendix \ref{app:z_dependence}. Moreover, the model assumes a universal value of $\eta$ across all haloes. In practice, this assumption may be oversimplified, as $\eta$ is likely to exhibit a dependence on the halo mass (e.g. more massive haloes, with deeper gravitational potential wells, should be able to retain the ejected gas more effectively).

  \subsection{Probes of baryons: kSZ and X-ray mass fractions} \label{ssec:model.barprobe}
    Here we will consider two gas probes that have been often used in the past to quantify the impact of baryonic feedback in the matter power spectrum: X-ray constraints on the bound gas fraction and measurements of the kinematic Sunyaev-Zel'dovich effect \citep[e.g.][]{Schneider:2018, Grandis:2023, Bigwood:2024, Hadzhiyska:2024}.

    \subsubsection{X-ray observations} \label{sssec:model.xrays}
      The gas bound within galaxy clusters is primarily composed of ionised hydrogen and metals, which emit photons in the X-ray regime. Hence, observations of X-ray radiation emanating from clusters enable the study of the bound gas component. The resulting measurements of the bound gas fraction can thus be used to constrain the BCM parameters $M_{\mathrm{c}}$ and $\beta$, governing the mass dependence of the bound fraction in Equation \eqref{eq:f_bgas}. To summarise the gas fractions measured in X-ray data, we parametrise the total mass of bound gas, $M_{\mathrm{bgas}}$, as in \citet{Grandis:2023},
      \begin{equation} \label{eq:M_bgas}
        \frac{M_{\mathrm{bgas}}}{M_{\mathrm{bgas}}^{\mathrm{piv}}} = A_{\mathrm{bgas}} \left(\frac{M_{500\mathrm{c}}}{M_{\mathrm{piv}}}\right)^{B_{\mathrm{bgas}}},
      \end{equation}
      where $M_{500\mathrm{c}}$ is the halo mass contained within a radius $r_{500\mathrm{c}}$, enclosing a total mass density corresponding to 500 times the critical density, $M_{\mathrm{piv}}$ is the median mass value of the cluster sample, and $M_{\mathrm{bgas}}^{\mathrm{piv}}$ is the mass normalisation at the pivot mass. The free parameters $A_{\mathrm{bgas}}$ and $B_{\mathrm{bgas}}$ describe the normalisation at the pivot mass and the power-law index of the gas mass to halo mass relation, respectively. Theoretical predictions for these parameters can be derived in terms of the bound gas fraction in a given halo, $f_{\mathrm{bgas}}$, which in X-ray studies is defined as
      \begin{equation} \label{eq:f_bgas_xrays}
        f_{\mathrm{bgas}} \equiv \frac{M_{\mathrm{bgas}}}{M_{500\mathrm{c}}}.
      \end{equation}
      Using the scaling relations from \citet{Grandis:2023}, the theoretical predictions for $A_{\mathrm{bgas}}$ and $B_{\mathrm{bgas}}$ are given by
      \begin{align} 
        A_{\mathrm{bgas}} &= \frac{M_{\mathrm{piv}}}{M_{\mathrm{bgas}}^{\mathrm{piv}}} \left. f_{\mathrm{bgas}} \right|_{M_{500\mathrm{c}} = M_{\mathrm{piv}}}, \label{eq:A_bgas} \\
        B_{\mathrm{bgas}} &= 1 + \left. \frac{\mathrm{d} \ln f_{\mathrm{bgas}}}{\mathrm{d} \ln M_{500\mathrm{c}}} \right|_{M_{500\mathrm{c}} = M_{\mathrm{piv}}}. \label{eq:B_bgas}
      \end{align}
      In this study, we use the parametrisation of the bound gas fraction implemented in \bemu, as given in Eq. \eqref{eq:f_bgas}. To generate our X-ray mock data for $A_{\mathrm{bgas}}$ and $B_{\mathrm{bgas}}$, we use the pivot masses and redshifts from \citet{Chiu:2018}, \citet{Chiu:2021}, and \citet{Akino:2021}, with $M_{\mathrm{bgas}}^{\mathrm{piv}} = 10^{13} M_{\mathrm{\odot}}$. Since observations are mostly sensitive to the X-ray signal originating from within $r_{500\mathrm{c}}$, X-ray cluster studies typically adopt the halo mass definition $M_{500\mathrm{c}}$. Hence, we use the concentration-mass relation from \citet{Ishiyama:2020} to convert from $M_{500\mathrm{c}}$ to $M_{200\mathrm{c}}$ to calculate the bound gas and stellar fractions, $f_{\mathrm{bgas}}$ and $f_*$, using Equations \eqref{eq:f_bgas} and \eqref{eq:f_*}.

\begin{figure}
        \centering
        \includegraphics[width=\linewidth]{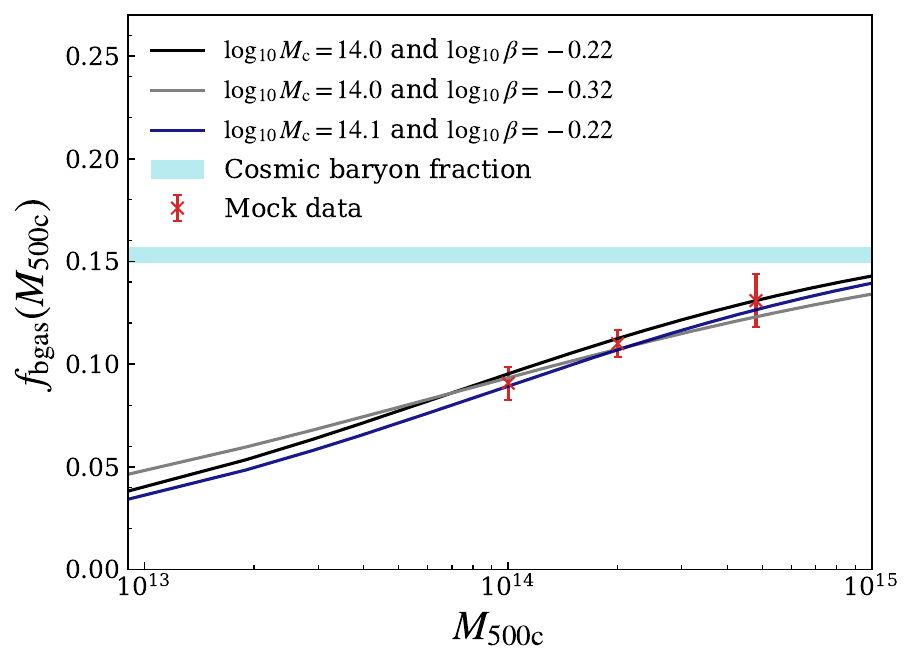}
        \caption{The bound gas fraction and its associated error from mock measurements of $A_{\mathrm{bgas}}$ and $B_{\mathrm{bgas}}$ based on the pivot masses and redshifts from recent X-ray studies, with $M_{\mathrm{bgas}}^{\mathrm{piv}} = 10^{13} M_{\odot}$. The error bars are derived from currently available measurements summarised in \citet{Grandis:2023}. The black curve shows the \bemu parametrisation for the bound gas fraction (Equation \ref{eq:f_bgas}) for the fiducial cosmology used to generate the mock measurements. The blue and grey curves show the sensitivity of the model to variations in $\log_{10} M_{\mathrm{c}}$ and $\log_{10} \beta$, respectively. The horizontal band shows the mean cosmological value of the bound gas fraction, defined as $f_{\mathrm{bgas}} = \Omega_{\mathrm{b}} / \Omega_{\mathrm{m}} - f_*$. The variation in the value of $f_{\mathrm{bgas}}$ across the band reflects differences in the stellar fraction, $f_*$, associated with each individual mock data point.}\label{fig:Xray_gas_fractions}
      \end{figure}

      We present the predictions for the bound gas fraction, $f_{\mathrm{bgas}}$, derived from the \bemu parametrisation in Equation \eqref{eq:f_bgas}, as a function of halo mass, $M_{500\mathrm{c}}$, in Fig. \ref{fig:Xray_gas_fractions}. To generate the mock data points, shown in red, we use a fixed set of BCM parameters to characterise the gas $(\log_{10} M_{\mathrm{c}}=14.0$, $\log_{10} \eta=-0.30$, $\log_{10} \beta=-0.22$, $\log_{10} M_1 =10.674$, $\log_{10} \theta_{\mathrm{inn}}=-0.86$, $\log_{10} \theta_{\mathrm{out}}=0.25$, $\log_{10} M_{\mathrm{inn}}=13.0)$. Here, all masses are in units of $M_{\odot}$. The blue and grey lines show the sensitivity of the model to variations in $\log_{10} M_{\mathrm{c}}$ and $\log_{10} \beta$, respectively. This illustrates the ability of X-ray gas fractions to constrain the bound gas abundance. Moreover, Fig. \ref{fig:Xray_gas_fractions} shows that smaller haloes exhibit a lower gas fraction compared to larger ones. This deficiency in gas is primarily attributed to the stronger influence of AGN feedback on smaller haloes, which are less capable of retaining gas due to their weaker gravitational field. The gas fraction approaches the mean cosmological value, $f_{\mathrm{bgas}} = \Omega_{\mathrm{b}} / \Omega_{\mathrm{m}} - f_*$, for haloes of larger mass, in agreement with hydrodynamical simulations \citep[e.g.][]{Sorini:2022, Ayromlou:2023}. 

      Subsequently, we use the mock data points for the bound gas fraction presented in Fig. \ref{fig:Xray_gas_fractions} to generate mock measurements for $A_{\mathrm{bgas}}$ and $B_{\mathrm{bgas}}$ via Equations \eqref{eq:A_bgas} and \eqref{eq:B_bgas}, respectively.

    \subsubsection{kSZ stacking observations} \label{sssec:model.ksz}
      The kSZ effect is a secondary anisotropy of the CMB in which a Doppler shift is imparted on CMB photons via inverse Compton scattering, which is induced by the bulk motion of ionised gas in galaxy clusters \citep{Sunyaev:1970,Sunyaev:1972}. In turn, this shifts the temperature of the CMB by
      \begin{equation} \label{eq:DeltaT}
        \frac{\Delta T_{\mathrm{kSZ}}}{T_{\mathrm{CMB}}} = - \sigma_{\mathrm{T}} \int \mathrm{d}l \, e^{-\tau(z)} n_{\mathrm{e}} \frac{\mathbf{v}_{\mathrm{e}} \cdot \hat{\mathbf{n}}}{c},
      \end{equation}
      where $\sigma_{\mathrm{T}}$ is the Thomson cross-section, $\tau(z)$ is the optical depth to Thomson scattering along the line-of-sight at redshift $z$, $n_{\mathrm{e}}$ is the physical number density of free electrons, $\mathbf{v}_{\mathrm{e}}$ is the peculiar velocity, and $c$ is the speed of light. The integral $\int \mathrm{d}l = \int \mathrm{d}\chi / (1+z)$ is taken along the line of sight in the direction given by the unit vector $\hat{\mathbf{n}}$. Since the kSZ signal is proportional to the integrated electron momentum along the line of sight, it can provide a direct measurement of the electron density, provided that the velocity field is known.

      We generate mock data for stacked kSZ measurements, in which the CMB map is correlated with the positions of galaxies weighted by their reconstructed velocities. The result may be interpreted as being proportional to the gas density profile. We assume measurements similar to those made by the Atacama Cosmology Telescope (ACT) DR5 and Planck \citep{Schaan:2021,Amodeo:2020} for the CMASS galaxy sample from the Baryon Oscillation Spectroscopic Survey (BOSS) \citep{BOSS:2013}, with an average halo mass of $M_{200\mathrm{c}} = 3 \times 10^{13} M_{\odot}$ and an average redshift of $z = 0.55$. Since the optical depth is below the percent level \citep{Planck:2016}, we can approximate the galaxy sample as optically thin, $e^{-\tau(z)} \approx 1$. In this limit, we can recast Equation \eqref{eq:DeltaT} in terms of the optical depth of the galaxy cluster, $\tau_{\mathrm{gal}}$, as
      \begin{equation} \label{eq:DeltaT_tau}
        \frac{\Delta T_{\mathrm{kSZ}}}{T_{\mathrm{CMB}}} = - \tau_{\mathrm{gal}} \frac{v_{\mathrm{e, r}}}{c},
      \end{equation}
      where $v_{\mathrm{e,r}}$ is the rms radial velocity of the free electrons, which we approximate as $v_{\mathrm{e, r}} = 1.06 \times 10^{-3} c$, following \citet{Schaan:2021} and \citet{Amodeo:2020}. Assuming spherical symmetry, the optical depth $\tau_{\mathrm{gal}}$ measured at an angular separation $\theta$ away from the centre of the cluster is given by
      \begin{equation} \label{eq:tau_gal}
        \tau_{\mathrm{gal}}(\theta) = 2 \sigma_{\mathrm{T}} \int_0^{r_{200\mathrm{c}}} \mathrm{d}l \, n_{\mathrm{e}}\left(\sqrt{d_A(z)^2 \theta^2 + l^2}\right),
      \end{equation}
      where $d_A(z)$ is the angular diameter distance, and the electron number density profile, $n_{\mathrm{e}}(r)$, is related to the total gas density profile, $\rho_{\mathrm{gas}}(r) = \rho_{\mathrm{bgas}}(r) + \rho_{\mathrm{egas}}(r)$, via
      \begin{equation} \label{eq:n_e_profile}
        n_{\mathrm{e}}(r) = \frac{X_{\mathrm{H}}+1}{2 m_{\mathrm{amu}}} \rho_{\mathrm{gas}}(r).
      \end{equation}
      Here, $X_{\mathrm{H}} = 0.76$ is the hydrogen mass fraction and $m_{\mathrm{amu}}$ is the atomic mass unit.

      It is important to note the potential limitations of this interpretation of the stacked kSZ profile. Specifically, we do not incorporate the 2-halo term, and the velocity reconstruction method has inherent shortcomings. Consequently, our interpretation of the constraining power of the kSZ is likely optimistic, and any real constraints derived from this analysis may be less stringent.

      Since the kSZ signal has the same frequency spectrum as the primary CMB, the CMB anisotropies themselves contaminate the kSZ map. To limit the estimator variance caused by large-scale CMB fluctuations, the standard stacked kSZ estimators apply a compensated aperture photometry filter (CAP) to the CMB map \citep{ACTPol:2015,Alonso:2016,Schaan:2021}. This method involves considering two concentric circles with radii $\theta_{\mathrm{d}}$ and $\sqrt{2} \theta_{\mathrm{d}}$, chosen such that the area enclosed by the inner disc and the outer ring are equal. Since the CMB fluctuations have a typical angular size much larger than $\theta_{\mathrm{d}}$, the CMB fluctuations will be approximately constant over the aperture. Therefore, the net CMB flux will be zero when the flux over the inner disc is subtracted from that over the outer ring, leaving only the kSZ flux. Mathematically, we can model the CAP filter using a window function,
      \begin{equation} \label{eq:cap_filter}
        W_{\theta_{\mathrm{d}}} = 
        \begin{cases}
          1 & \text{for } 0 < \theta < \theta_{\mathrm{d}}, \\
          -1 & \text{for } \theta_{\mathrm{d}} < \theta < \sqrt{2} \theta_{\mathrm{d}}, \\
          0 & \text{otherwise}.
        \end{cases}
      \end{equation}
      Hence, the observed kSZ signal is a convolution of Equation \eqref{eq:DeltaT_tau} with the CAP filter in Equation \eqref{eq:cap_filter} and the CMB beam function. For the latter, we approximate the beam as a Gaussian with a full-width at half-maximum (FWHM) of 2.1 arcminutes, corresponding to a diffraction-limited telescope with a $\sim6\,{\rm m}$ diameter aperture at 98 GHz \citep{Schaan:2021}. To generate mock measurements of the stacked kSZ profile, we use 30 bins for $\theta_{\mathrm{d}}$, ranging from $\theta_{\mathrm{d}} = 0.5$ arcmin to $6.0$ arcmin.

      From Equation \eqref{eq:n_e_profile}, we observe that the kSZ signal is directly sensitive to the total gas density profile, with the bound and ejected gas profiles modelled by Equations \eqref{eq:rho_bgas} and \eqref{eq:rho_egas}, respectively. Thus, tracing the kSZ signal provides a means to constrain the baryonic feedback parameters that describe both the bound and ejected gas profiles. In contrast to X-ray data alone, which is mostly sensitive to the bound gas component, kSZ data allows us to constrain the parameter $\eta$, which governs the physical extent of the ejected gas. However, it is important to note that, in practise, our mock measurements of the kSZ profile cannot provide individual constraints on the bound gas parameters, $M_{\mathrm{c}}$ and $\beta$. This limitation arises because our kSZ data probes only a single halo mass and redshift, leading to a degeneracy between $M_{\mathrm{c}}$ and $\beta$ that cannot be broken without additional data.

\begin{figure}
        \centering
        \includegraphics[width=\linewidth]{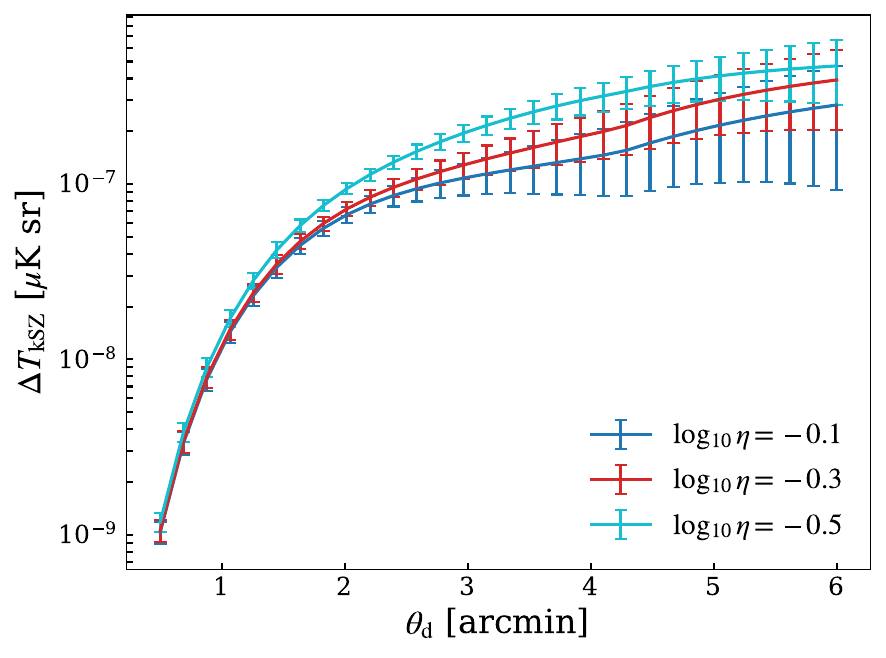}
        \caption{The CMB temperature shift due to the kSZ effect as a function of the aperture radius $\theta_{\mathrm{d}}$ of the CAP filter and for different feedback strengths. The model profile is convolved with the ACT DR5 f90 beam profile with FWHM = 2.1 arcmin. The error bars are derived from the baseline noise level of the Simons Observatory.}
        \label{fig:DeltaT_ksz_vs_theta_d}
    \end{figure}

      In Fig. \ref{fig:DeltaT_ksz_vs_theta_d}, we illustrate the kSZ signal as a function of the CAP aperture radius for a halo of fixed mass. In this figure, we explore the impact of varying the strength of baryonic feedback by modifying the parameter $\log_{10} \eta$. We find that an increase in $\log_{10} \eta$ decreases in the kSZ signal at small aperture radii $\theta_{\mathrm{d}}$, as a larger fraction of the gas is ejected beyond the virial radius. Independent of the feedback strength, the kSZ signal asymptotically approaches the same value at large $\theta_{\mathrm{d}}$, since the total mass enclosed within the halo remains constant.

      It is worth emphasising the limitations of our treatment of the kSZ data in this analysis. On the one hand, we neglect a number of potentially important sources of uncertainty that could degrade the constraining power of kSZ data. First, the amplitude of the stacked kSZ signal depends on the quality of the reconstructed galaxy velocities. Uncertainties in the amplitude of these velocities (e.g. due to uncertainties in galaxy bias, or inaccuracies in the reconstruction algorithm), should be propagated when using the kSZ measurements to extract information about the gas distribution. Secondly, we will interpret the stacked kSZ measurements as being solely sensitive to the electron density profile of haloes of a given mass. This neglects important additional contributions, including the ``2-halo'' contribution from surrounding structures, the impact of miscentering and satellite galaxies, and additional correlations between the velocity and density fields. Specifically, the presence of satellite galaxies and the effects of miscentering can introduce a bias in the measured kSZ amplitude. To illustrate, \citet{McCarthy:2024} demonstrate that accounting for satellite contributions can enhance the kSZ signal by approximately $20\%$. On the other hand, we have only considered a single galaxy sample, probing a single halo mass, at a single effective redshift. kSZ measurements are able to probe a range of redshifts and masses \citep{2407.07152,2503.19870}, which could allow us to constrain $\log_{10}M_c$ and $\beta$ separately, as well as the potential mass/redshift dependence of $\eta$. Our analysis should, nevertheless, provide an indication of whether the sensitivity of current and future CMB observations will be enough to constrain feedback parameters at the rough level required of Stage-IV weak lensing data.

  \subsection{Likelihood} \label{ssec:model.like}
    \begin{table*}
      \centering
      \begin{tabular}{llrr} 
        \hline
        Parameter & Description & Fiducial Value & Prior \\
        \hline
        \textbf{Cosmological} &&& \\
        $\Omega_{\mathrm{m}}$ & density of matter in units of the critical density of the Universe & $0.30967$ & $\mathcal{N}(0.30967, 0.1)$ \\
        $\sigma_8$ & cold mass linear mass variance in $8 h^{-1} \text{Mpc}$ spheres & $0.8102$ & $\mathcal{N}(0.8102, 0.1)$ \\
        $\Omega_{\mathrm{b}}$ & density of baryons in units of the critical density of the Universe & $0.04897$ & $-$ \\
        $h$ & dimensionless Hubble constant & $0.6766$ & $\mathcal{N}(0.6766, 0.1)$ \\
        $n_{\mathrm{s}}$ & scalar spectral index & $0.9665$ & $\mathcal{N}(0.9665, 0.1)$ \\
        $\Sigma m_\nu$ & sum of neutrino masses in units of eV & $0.06$ & $\mathcal{N}(0.06, 0.1)$ \\ 
        \hline
        \textbf{Baryonic Feedback} &&& \\
        $\log_{10} M_{\mathrm{c}}$ & characteristic halo mass for which half the gas is retained & $14.0$ & $\mathrm{U}[9.0, 15.0]$ \\
        $\log_{10} \eta$ & directly proportional to the radius of ejected gas from the halo & $-0.3$ & $\mathrm{U}[-0.7, 0.7]$ \\
        $\log_{10} \beta$ & describes the rate at which the depletion of gas increases towards smaller haloes & $-0.22$ & $\mathrm{U}[-1.0, 0.7]$ \\
        $\log_{10} M_1$ & characteristic mass of haloes that host a central galaxy at $z=0$ & $10.674$ & $\mathrm{U}[9.0, 13.0]$ \\
        $\log_{10} M_{\mathrm{inn}}$ & characteristic transition mass of the density profile of hot gas in haloes & $13.0$ & $\mathrm{U}[9.0, 13.5]$ \\
        $\log_{10} \theta_{\mathrm{inn}}$ & controls the inner radius of the bound gas density profile & $-0.86$ & $\mathrm{U}[-2.0, 0.5]$ \\
        $\log_{10} \theta_{\mathrm{out}}$ & controls the outer radius of the bound gas density profile & $0.25$ & $\mathrm{U}[0.0, 0.5]$\\
        \hline
        \textbf{Intrinsic Alignments} &&& \\
        $A_{\mathrm{IA},0}$ & amplitude of the NLA model for IAs & $0.0$ & $\mathcal{N}(0, 1.0)$\\
        $\eta_{\mathrm{IA}}$ & slope of the NLA model for IAs & $0.0$ & $\mathcal{N}(0, 1.0)$\\
        \hline
      \end{tabular}
      \caption{Qualitative descriptions of cosmological parameters, baryonic parameters used in the \bemu model, and intrinsic alignment parameters. Note that all masses are in units of $M_{\odot}$ throughout this work. Here, fiducial value refers to the parameter value used to generate the mock data. We use Gaussian priors, centred on the corresponding fiducial value, for the cosmological and intrinsic alignment parameters. We use the flat \bemu priors for the baryonic parameters. We note that the adopted priors on the cosmological parameters ensure they remain within the valid calibration range of the \bemu emulator.}
      \label{tab:free_parameters}
    \end{table*}

    Throughout this study, we adopt a {\sl Planck} cosmology \citep{1807.06209} with parameter values $\{\Omega_{\mathrm{c}}, \Omega_{\mathrm{b}}, h, n_{\mathrm{s}}, \sigma_8, \Sigma m_{\nu}\} =$ $\{0.2607$, $0.04897$, $0.6766$, $0.9665$, $0.8102, 0.06\}$ as the fiducial model to generate the mock data. We assume that the mock data, $\mathbf{d}$, follows a Gaussian likelihood with the $\chi^2$ statistic for the set of parameters, $\Vec{\theta}$, given by
    \begin{equation}
      \chi^2 = - 2 \log p(\mathbf{d}|\Vec{\theta}) = (\mathbf{d} - \mathbf{t}(\Vec{\theta}))^{\mathrm{T}} \mathsf{C}^{-1} (\mathbf{d} - \mathbf{t}(\Vec{\theta})) + K,
    \end{equation}
    where $\mathbf{t}$ is the theoretical prediction for the data, $\mathsf{C}$ is the covariance matrix of the data, and $K$ is a normalisation constant. To sample the posterior distribution, we use \texttt{Cobaya} \citep{Cobaya}, which implements the Metropolis-Hastings Markov Chain Monte Carlo (MCMC) method \citep{Metropolis:1953}. We simultaneously marginalise over five cosmological parameters $(\Omega_{\mathrm{m}}, \sigma_8, h, n_{\mathrm{s}}, \Sigma m_{\nu})$, two intrinsic alignment parameters $(A_{\mathrm{IA,0}}, \eta_{\mathrm{IA}})$, and the seven baryonic parameters implemented in \bemu $(\log_{10} M_{\mathrm{c}}$, $\log_{10} \eta$, $\log_{10} \beta$, $\log_{10} M_1$, $\log_{10} \theta_{\mathrm{inn}}$, $\log_{10} \theta_{\mathrm{out}}$, $\log_{10} M_{\mathrm{inn}})$. The calibratable systematics, $\Delta z_i$ and $m_i$, are marginalised over using the analytical approximation outlined in Section \ref{sssec:model.photo_z_and_multbias}, which avoids introducing an additional two free parameters for each redshift bin into the likelihood function. A summary of the free parameters and their priors is presented in Table \ref{tab:free_parameters}.

   We justify our selection of parameters as follows. Although the primordial spectral index, $n_{\rm s}$, is tightly constrained by measurements of the CMB, we marginalise over it in this analysis to incorporate complementary constraints from other cosmological probes. In contrast, we fix the cosmological baryon fraction, $\Omega_{\rm b}$, given its precise determination from both the CMB and Big Bang Nucleosynthesis (BBN). This approach allows for an independent cross-validation of the CMB-derived baryon fraction through BBN, making additional constraints from weak lensing non-essential in this context.

    We use CCL \citep{CCL} to generate LSST-like weak lensing data for multipoles up to $\ell_{\mathrm{max}} = 2000$. To account for the finite width of the $\ell$ bins, we use a linear spacing of $\Delta \ell = 10$ up to a given value $\ell_{\mathrm{linear}}$. Beyond this value, we use logarithmic sampling with ten $\ell$ values per decade. The value of $\ell_{\mathrm{linear}}$ is chosen such that the separation between adjacent $\ell$ values beyond $\ell_{\mathrm{linear}}$ using logarithmic sampling is greater than or equal to $\Delta \ell_{\mathrm{linear}} = 10$. We construct the covariance matrix for the LSST cosmic shear power spectra using the Knox formula,
    \begin{equation}
      \mathrm{Cov}(C_\ell^{ij}, C_{\ell'}^{km}) = \frac{\delta_{\ell\ell'}}{f_{\mathrm{sky}}(2\ell+1)} \left(C_\ell^{ik} C_\ell^{jm} + C_\ell^{im} C_\ell^{jk}\right),
    \end{equation}
    where $f_{\mathrm{sky}} = 0.4$ is the fraction of the sky covered by the weak lensing survey. We also include the correction ${\sf T}\,{\sf P}\,{\sf T}^T$ from marginalising analytically over photometric redshift uncertainties and multiplicative shape biases (see Equation \eqref{eq:updated_cov}). 

    It is important to note that, in principle, the shear dataset for the LSST 10-year sample described in Section \ref{ssec:model.cshear} is sensitive to  redshifts up to $z=4.0$, and physical wavenumbers up to $k = 10 \, \mathrm{Mpc}^{-1}$, whereas the \bemu emulator is calibrated only up to redshift $z=1.5$ and wavenumbers of $k = 5 \, h \mathrm{Mpc}^{-1}$, which corresponds to $k \sim 3.5 \, \mathrm{Mpc}^{-1}$ in our fiducial cosmology. However, the contribution from shape noise to the power spectrum covariance limits the sensitivity to the smallest angular scales, and the lensing kernels peak half-way between the observer and the source, suppressing the contribution from structure at high redshifs. This allows us to rely on relatively simple extrapolation schemes for $\bemu$. These are described in Appendix \ref{app:bacco_implementation}, where we also assess their validity for the dataset simulated here.

    For the X-ray mock data, we assume that the different cluster samples exhibit minimal to no overlap, and therefore we treat the measurements as mutually independent. As a result, the covariance matrix for the X-ray mock data takes a diagonal form with the errors for the $A_{\mathrm{bgas}}$ and $B_{\mathrm{bgas}}$ measurements in each cluster sample on the leading diagonal. We use the errors obtained from near-term X-ray samples, as given in \citet{Chiu:2018}, \citet{Chiu:2021}, and \citet{Akino:2021}. We also consider potential futuristic data based on the eROSITA survey, as discussed in Section \ref{ssec:results.xrayksz}.

    The covariance for the kSZ mock data can be calculated analytically, as described in \citet{Alonso:2016}:
    \begin{align}
      \mathrm{Cov}(\Delta T_{\mathrm{kSZ}}(\theta_1), \Delta T_{\mathrm{kSZ}}(\theta_2)) &= 2\pi \theta_1^2 \theta_2^2 \, \times \nonumber \\
      &\int \mathrm{d}\ell \, \ell \, C_N(\ell) \widetilde{W}(\ell | \theta_1) \widetilde{W}(\ell | \theta_2), \label{eq:ksz_covariance}
    \end{align}
    where $C_N(\ell)$ is the noise power spectrum which includes contributions from both the CMB and instrument noise, $C_N(\ell) = C_\ell^{\mathrm{CMB}} + N_\ell$, and $\widetilde{W}$ is the Fourier transform of the CAP filter given in Equation \eqref{eq:cap_filter},
    \begin{equation}
      \widetilde{W}(\ell | \theta_{\mathrm{d}}) = \frac{2 J_1(\ell \theta_{\mathrm{d}}) - \sqrt{2} J_1(\sqrt{2}\ell \theta_{\mathrm{d}})}{\ell \theta_{\mathrm{d}}},
    \end{equation}
    where $J_1(x)$ is the first-order cylindrical Bessel function. As described in Section \ref{ssec:results.xrayksz}, we will consider future kSZ measurements based on the sensitivities of the Simons Observatory and CMB-S4 experiments \citep{SimonsObservatory:2018,CMB-S4:2024}.


\section{Results} \label{sec:results}
  Using the data and theory pipeline discussed in the previous sections, we now perform an inference analysis using the likelihood and MCMC sampling method described in Section \ref{ssec:model.like}. We first quantify the level of information loss in cosmic shear analyses due to lack of understanding of baryonic effects. Then, we determine the level to which different baryonic parameters must be calibrated to avoid this information loss, and the ability of X-ray and kSZ data to achieve this requirement. Finally, we examine the cosmological constraints that may be obtained by combining cosmic shear with forecasts of X-ray and kSZ data as external calibrators.

  \subsection{Information loss due to baryonic effects} \label{ssec:results.info_loss}
    As a first result, we quantify the extent to which having freedom over the baryonic parameters degrades the cosmological constraints for LSST-like cosmic shear data. We use the complete mock dataset up to multipoles $\ell=2000$ in order to forecast the constraints achievable with the full LSST dataset in the future. Consequently, we do not evaluate the impact of applying scale cuts in this analysis.

\begin{figure}
    \centering
    \includegraphics[width=\linewidth]{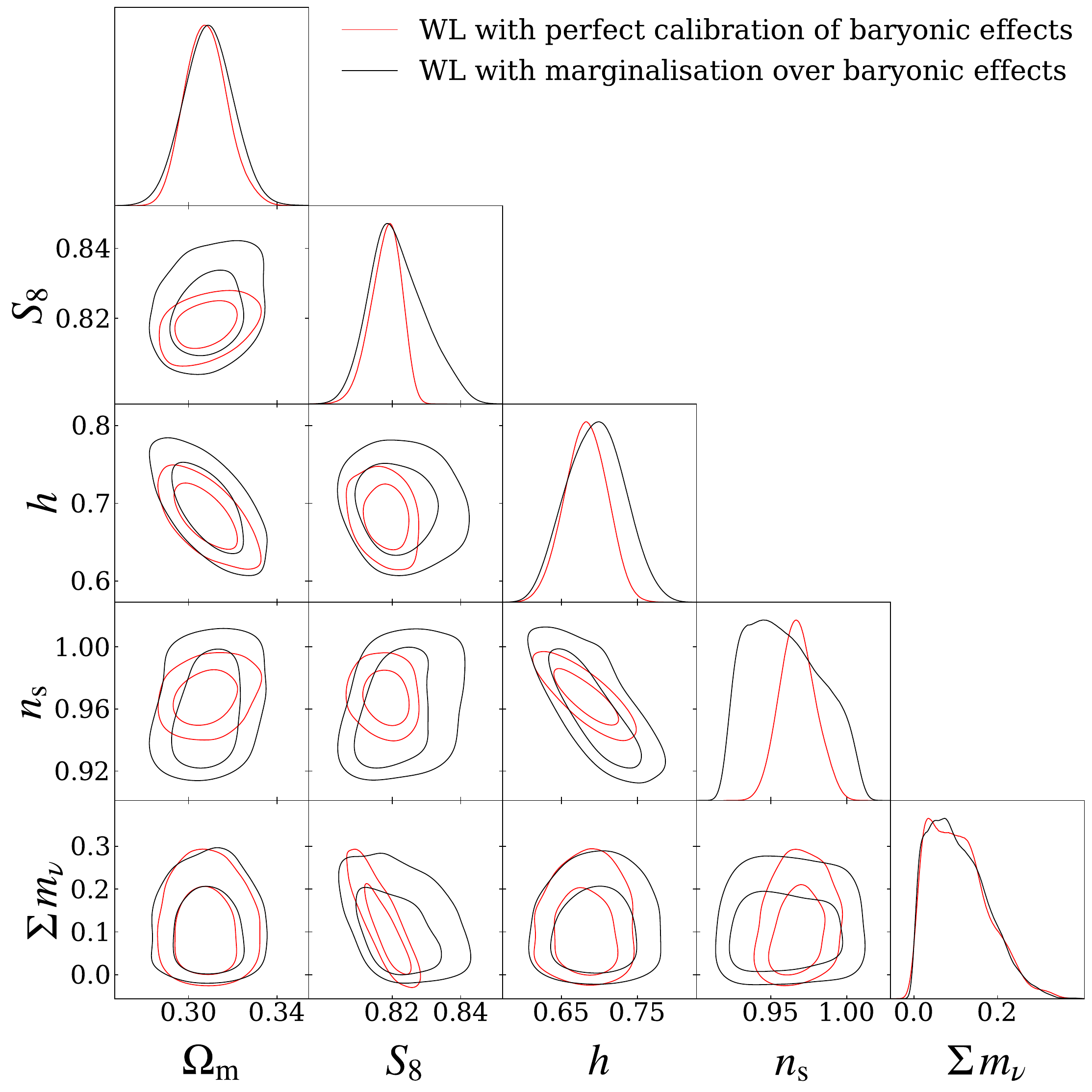}
    \caption{The marginalised posteriors for $\Omega_{\mathrm{m}}$, $S_8 = \sigma_8 \sqrt{\Omega_{\mathrm{m}}/0.3}$, $h$, $n_{\mathrm{s}}$, and $\Sigma m_\nu$ obtained from LSST-like weak lensing data up to multipoles of $\ell = 2000$. We compare the posteriors obtained under the marginalisation over baryonic effects (black) to the case where the baryonic parameters are kept fixed at the values used to generate the mock data (red). The inner and outer contours show the 95$\%$ and $68\%$ confidence levels, respectively. We marginalise over intrinsic alignments, photometric redshift uncertainties, and multiplicative shape biases in both cases.}
    \label{fig:info_loss_baryons}
  \end{figure}

    In Fig. \ref{fig:info_loss_baryons}, the posterior constraints obtained assuming perfect knowledge of baryonic  effects (red) are compared to those obtained after marginalising over baryonic feedback parameters (black). The numerical constraints are listed in the first two rows of Table \ref{tab:cosmo_errors}. We find that the errors on $S_8$, the parameter best determined by current weak lensing surveys, increase by a factor of 1.9 when marginalising over baryonic parameters. Future weak lensing data will be able to make precision measurements of other cosmological parameters, which are also affected by uncertainties on the impact of baryons. In particular, we find that the errors on the Hubble parameter, $h$, and the scalar spectral index, $n_{\mathrm{s}}$, grow by factors of 1.4 and 1.8, respectively. This highlights the importance of self-calibrating baryonic effects in order to fully benefit from the cosmological power of the upcoming Stage-IV weak lensing data. Furthermore, we find that, interestingly, baryonic effects do not affect the constraints on neutrino masses that are obtained from weak lensing data alone. This may change in the presence of additional cosmological data (e.g. from CMB primary anisotropies, or by combining weak lensing and galaxy clustering), but we leave this study for future work. We note that the loss of constraining power in $S_8$ comes mainly from $\sigma_8$ rather than from $\Omega_{\rm m}$. This is because the amplitude of the lensing power spectrum is influenced by both the intrinsic amplitude of the matter power spectrum and the distance-redshift relation. Baryonic effects affect the amplitude of the matter power spectrum, and not the distance-redshift relation, which results in a direct impact on $\sigma_8$ but not $\Omega_{\rm m}$.

  \begin{table*}
    \centering
    \begin{tabular}{lcccc}
        \hline
        Tracer(s) & $S_8$ & $\Omega_{\mathrm{m}}$ & $h$ & $n_{\mathrm{s}}$ \\
        \hline
        WL only with baryons fixed & $0.8181^{+0.0081}_{-0.0090}$ & $0.308^{+0.019}_{-0.018}$ & $0.682^{+0.051}_{-0.055}$ & $0.968^{+0.024}_{-0.022} $ \\
        & & & & \\
        WL only & $0.8220^{+0.0170}_{-0.0140}$ & $0.309^{+0.022}_{-0.021}$ & $0.695^{+0.073}_{-0.071}$ & $0.959^{+0.043}_{-0.038}$ \\
        & & & & \\
        WL + near-term X-ray & $0.8210^{+0.0150}_{-0.0130}$ & $0.309^{+0.019}_{-0.018}$ & $0.693^{+0.071}_{-0.068}$ & $0.962^{+0.041}_{-0.039}$ \\
        & & & & \\
        WL + near-term kSZ & $0.8200^{+0.0090}_{-0.0100}$ & $0.309^{+0.015}_{-0.016}$ & $0.678^{+0.049}_{-0.040}$ & $0.971^{+0.035}_{-0.039}$ \\
        & & & & \\
        WL + near-term X-ray + near-term kSZ & $0.8196^{+0.0094}_{-0.0100}$ & $0.308^{+0.015}_{-0.016}$ & $0.683^{+0.054}_{-0.052}$ & $0.970^{+0.035}_{-0.037}$ \\
        & & & & \\
        WL + long-term X-ray & $0.8210^{+0.0130}_{-0.0130}$ & $0.309^{+0.014}_{-0.015}$ & $0.688^{+0.066}_{-0.061}$ & $0.963^{+0.042}_{-0.040}$ \\
        & & & & \\
        WL + long-term kSZ & $0.8196^{+0.0087}_{-0.0098}$ & $0.308^{+0.014}_{-0.016}$ & $0.683^{+0.054}_{-0.034}$ & $0.968^{+0.034}_{-0.037}$ \\
        & & & & \\
        WL + long-term X-ray + long-term kSZ & $0.8196^{+0.0088}_{-0.0096}$ & $0.309^{+0.011}_{-0.012}$ & $0.679^{+0.032}_{-0.027}$ & $0.972^{+0.024}_{-0.024}$ \\
        & & & & \\
        WL + near-term X-ray + near-term kSZ with $n_{\rm gal}\times10$ & $0.8196^{+0.0078}_{-0.0086}$ & $0.309^{+0.014}_{-0.014}$ & $0.672^{+0.040}_{-0.028}$ & $0.974^{+0.027}_{-0.032}$ \\
        \hline
    \end{tabular}
    \caption{The constraints on $S_8$, $\Omega_{\mathrm{m}}$, $h$, and $n_{\mathrm{s}}$ obtained from different combinations of large-scale structure tracers. We report the mean marginal value of each parameter and their associated errors given by the 95$\%$ confidence level. Here, WL refers to LSST-like weak lensing data. We marginalise over intrinsic alignments, photometric redshift uncertainties, and multiplicative shape biases for the weak lensing data. In all cases other than the first row, we marginalise over the parameters describing baryonic effects. Note that we do not report the constraints on the neutrino mass, as weak lensing is not able to place tight constraints on this parameter.}
    \label{tab:cosmo_errors}
  \end{table*}

\begin{figure}
    \centering
    \includegraphics[width=\linewidth]{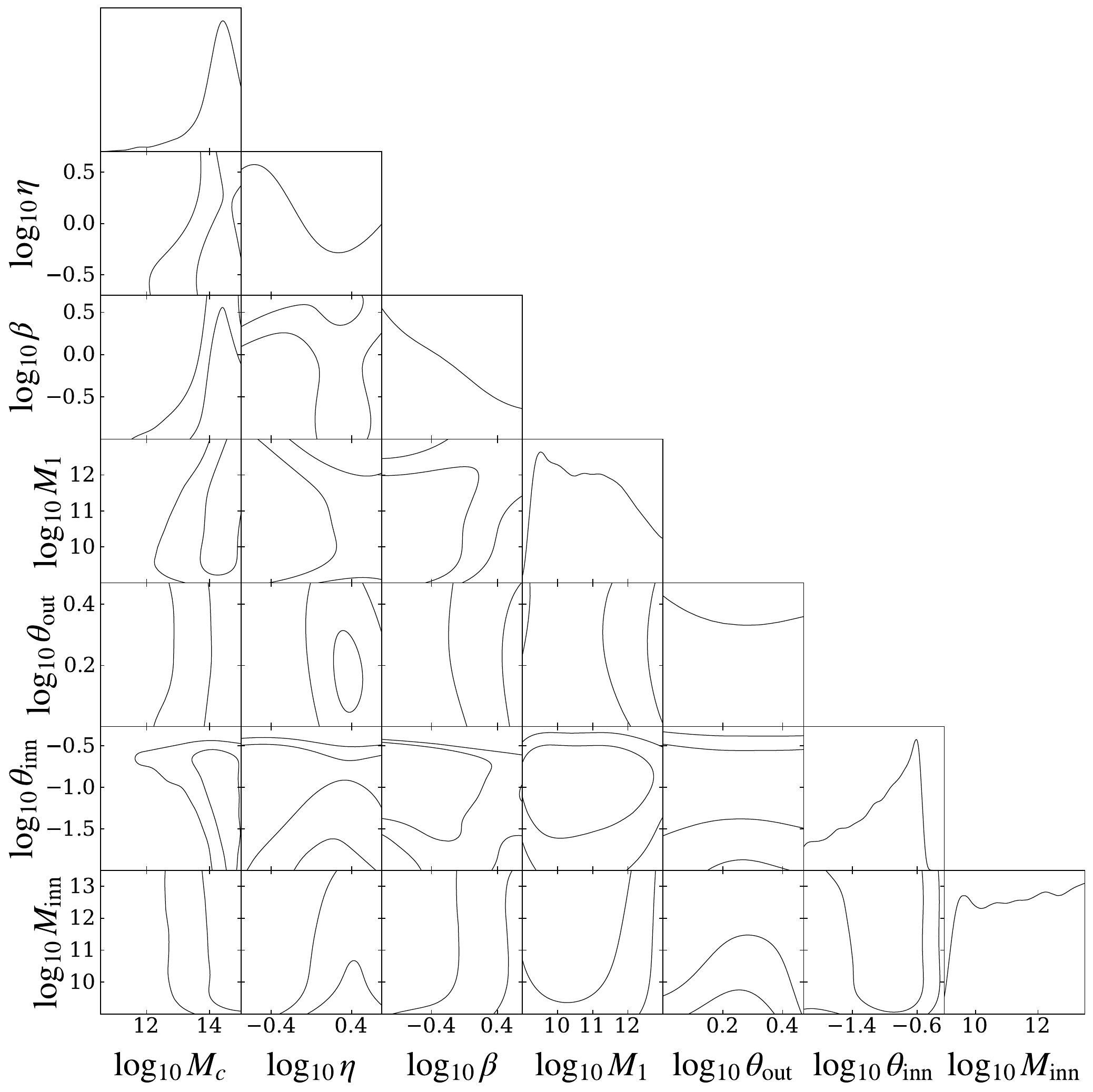}
    \caption{The marginalised posteriors on the baryonic parameters for LSST-like cosmic shear data alone. The inner and outer contours show the 68$\%$ and $95\%$ confidence levels, respectively. We marginalise over intrinsic alignments, photometric redshift uncertainties, and multiplicative shape biases.}
    \label{fig:wl_baryon_constraints}
  \end{figure}

  It is also interesting to study the level to which future, more sensitive, cosmic shear data will be able to self-calibrate baryonic effects, building on the work conducted in \citet{Preston:2024}. Fig. \ref{fig:wl_baryon_constraints} shows the marginalised posteriors on the baryonic parameters obtained from this analysis. The only baryonic parameter for which meaningful constraints can be obtained is $\log_{10}M_{\mathrm{c}}$. We note that the upper bound on $M_{\mathrm{c}}$ is due to the upper limit of the \bemu prior. This parameter determines the abundance of bound gas in haloes (and hence also the fraction of ejected gas), and thus controls the amplitude of the baryonic power suppression. This result is in agreement with the results found in the literature when analysing current cosmic shear datasets \citep{Arico:2023, Garcia-Garcia:2024}, and it is interesting to observe that the enhanced sensitivity of LSST cosmic shear data will not allow it to significantly constrain other baryonic parameters on its own. As we show in the next section, although cosmic shear is only able to constrain $\log_{10}M_{\mathrm{c}}$, other baryonic parameters play a significant role in significantly degrading cosmological constraints. We note, however, that higher-order statistics, such as the bispectrum, are expected to enhance the WL-only constraints on baryonic parameters \citep{Foreman:2019, Arico:2020bispectrum}.

  \subsection{Calibration requirements for baryonic effects}\label{ssec:results.calreq}
    As a next step, we identify the baryonic parameters that most significantly contribute to the degradation of the cosmological constraints, and estimate the level to which they must be calibrated to minimise this degradation. To achieve this, we perform a series of MCMC analyses in which one baryonic parameter is varied at a time, keeping all other parameters fixed. This approach allows us to isolate the individual impact of each baryonic parameter on the cosmological constraints. Note that this approach will not capture the internal degeneracies between different baryonic parameters, which usually lead to further degradation in the cosmological constraints. Nevertheless, it will allow us to obtain a rough estimate of the calibration requirements.

\begin{figure}
      \centering
      \includegraphics[width=\linewidth]{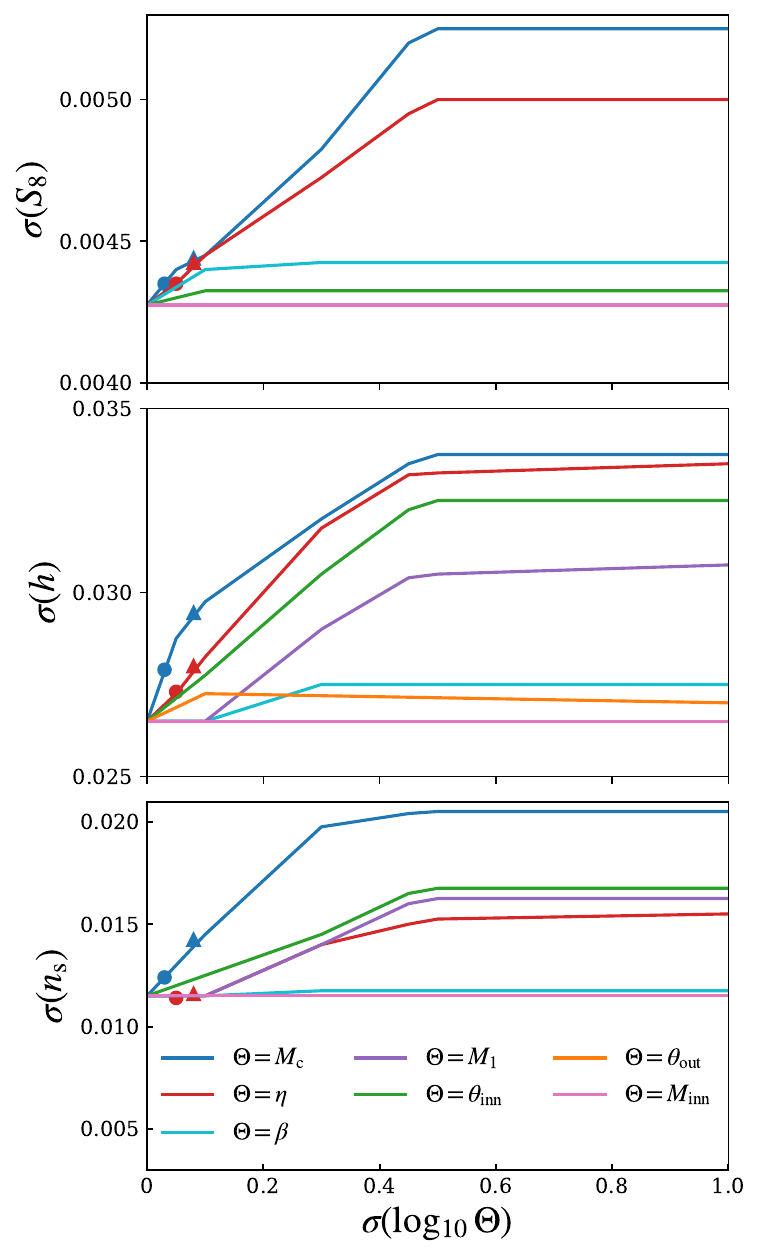}
      \caption{The errors on $S_8$, $h$, and $n_{\mathrm{s}}$, denoted as $\sigma(S_8)$, $\sigma(h)$, and $\sigma(n_{\mathrm{s}})$, respectively, as a function of the error on each baryonic parameter, denoted as $\sigma(\log_{10} \Theta)$ where $\Theta = \{M_{\mathrm{c}}, \eta, \beta, M_1, \theta_{\mathrm{inn}}, \theta_{\mathrm{out}}, M_{\mathrm{inn}}\}$. Here, the triangular and circular points represent the level to which we can constrain each parameter using near-term and long-term data from external tracers of the gas distribution in and around haloes, respectively. These constraints come from X-ray and kSZ data for $M_{\mathrm{c}}$ and $\eta$, respectively.}
      \label{fig:errors_on_cosmo_params}
    \end{figure}

    We focus on the constraints obtained on $S_8$, $h$, and $n_{\mathrm{s}}$, as these are the parameters most affected by uncertainties on baryonic effects (see Fig. \ref{fig:info_loss_baryons}). In Fig. \ref{fig:errors_on_cosmo_params}, we present the errors on these parameters, as a function of the calibration prior in each of the baryonic parameters. Our results indicate that variations in $\log_{10} M_{\mathrm{c}}$ and $\log_{10} \eta$ most substantially degrade the constraints on $S_8$, while the influence of the remaining baryonic parameters is negligible. Quantitatively, the error on $S_8$ grows roughly linearly with the uncertainty on $\log_{10} M_{\mathrm{c}}$ and $\log_{10} \eta$, and both parameters must be calibrated at the level of $\sigma(\log_{10} M_{\mathrm{c}}) \sim \sigma(\log_{10} \eta) \lesssim 0.1$ in order to avoid degrading the constraints on $S_8$ by more than $\sim 10\%$. When expressed in terms of the fractional uncertainty on the value of $M_{\mathrm{c}}$ and $\eta$ themselves, this corresponds to a $\sim 10-20\%$ calibration requirement\footnote{The fractional uncertainty in a given quantity corresponds to the uncertainty in its natural logarithm, as given by $\Delta \ln(x) = \Delta x/x$. When expressed in terms of base-10 logarithms, this relation becomes $\Delta x/x = \ln{(10)} \times \Delta \log_{10} (x) \approx 2.3\,\Delta \log_{10} (x)$.}.

    The two bottom panels of Fig. \ref{fig:errors_on_cosmo_params} show the constraints on $h$ and $n_s$. In both cases, as for $S_8$, the errors on the cosmological parameters are most affected by the uncertainties in $\log_{10}M_{\mathrm{c}}$, with $\log_{10}\eta$ also playing a significant role (although less so in the case of $n_{\mathrm{s}}$). Interestingly, however, we find that marginalising over the inner scale radius, $\log_{10} \theta_{\mathrm{inn}}$, and the stellar parameter, $\log_{10} M_1$, can significantly degrade the constraints on both $h$ and $n_{\mathrm{s}}$, even if this does not propagate to $S_8$. However, as we will see in the next section, neither of the external gas probes considered here is sufficiently sensitive to these parameters.

  \subsection{External calibrators for baryonic effects}\label{ssec:results.xrayksz}
    We can use external tracers of the gas distribution in and around haloes to self-calibrate the baryonic parameters, which in turn allows for an improvement in the cosmological constraints extracted from weak lensing analyses. Here, we explore the combination of X-ray gas fraction estimates and stacked kSZ measurements, and determine whether this data will be able to match the calibration requirements described in the previous section. We will consider two types of datasets. First, we will use mock observations compatible with data that already exists or will become available in the near future, certainly within the lifetime of LSST and other Stage-IV weak lensing surveys. We will also consider more futuristic data, based on larger X-ray samples or more sensitive CMB observations. We will refer to these mock datasets as the ``near-term'' and ``long-term'' external calibrators.

    For the near-term X-ray dataset, we generate mock gas fraction measurements assuming a sample similar to the currently available data used in \citet{Grandis:2023}, corresponding to a total of 540 galaxy clusters from SPT, HSC-XXL, and eFEDS. We use the measurement uncertainties of \citet{Chiu:2018,Chiu:2021,Akino:2021}, on which the \citet{Grandis:2023} data is based. For near-term kSZ measurements, we assume noise levels that are compatible with what could be achieved by the Simons Observatory (SO) \citep{SimonsObservatory:2018}. Specifically, we use the official temperature noise curves presented in \citet{SimonsObservatory:2018} for the ``baseline'' noise level of the SO Large Aperture Telescope (LAT), assuming a $2.1\,{\rm arcmin}$ FWHM Gaussian beam. The resulting covariance matrix is estimated using Equation \eqref{eq:ksz_covariance}, including contributions from both instrumental noise and primary CMB fluctuations. For the long-term X-ray dataset, we construct mock observations based on the full eROSITA sample, which includes 5,259 galaxy clusters \citep{Ghirardini:2024}. As this represents approximately an order of magnitude increase in cluster count compared to the near-term mock dataset, we estimate the covariance matrix for the long-term X-ray measurements by scaling the uncertainties of the near-term dataset by a factor of $\sqrt{10}$. For the long-term kSZ dataset, we use the noise power spectrum for the CMB-S4 experiment to generate the covariance matrix. Specifically, we assume the noise parameters listed in Table 1 of \citet{CMB-S4:2024} for the 90 GHz channel. However, we note that the CMB-S4 mission has unfortunately been discontinued since conducting this work. At present, the next phase of CMB observations remains uncertain. One possibility is the CMB-HD or future extensions of the Simons Observatory \citep{CMB-HD}.

\begin{figure}
      \centering
      \includegraphics[width=\linewidth]{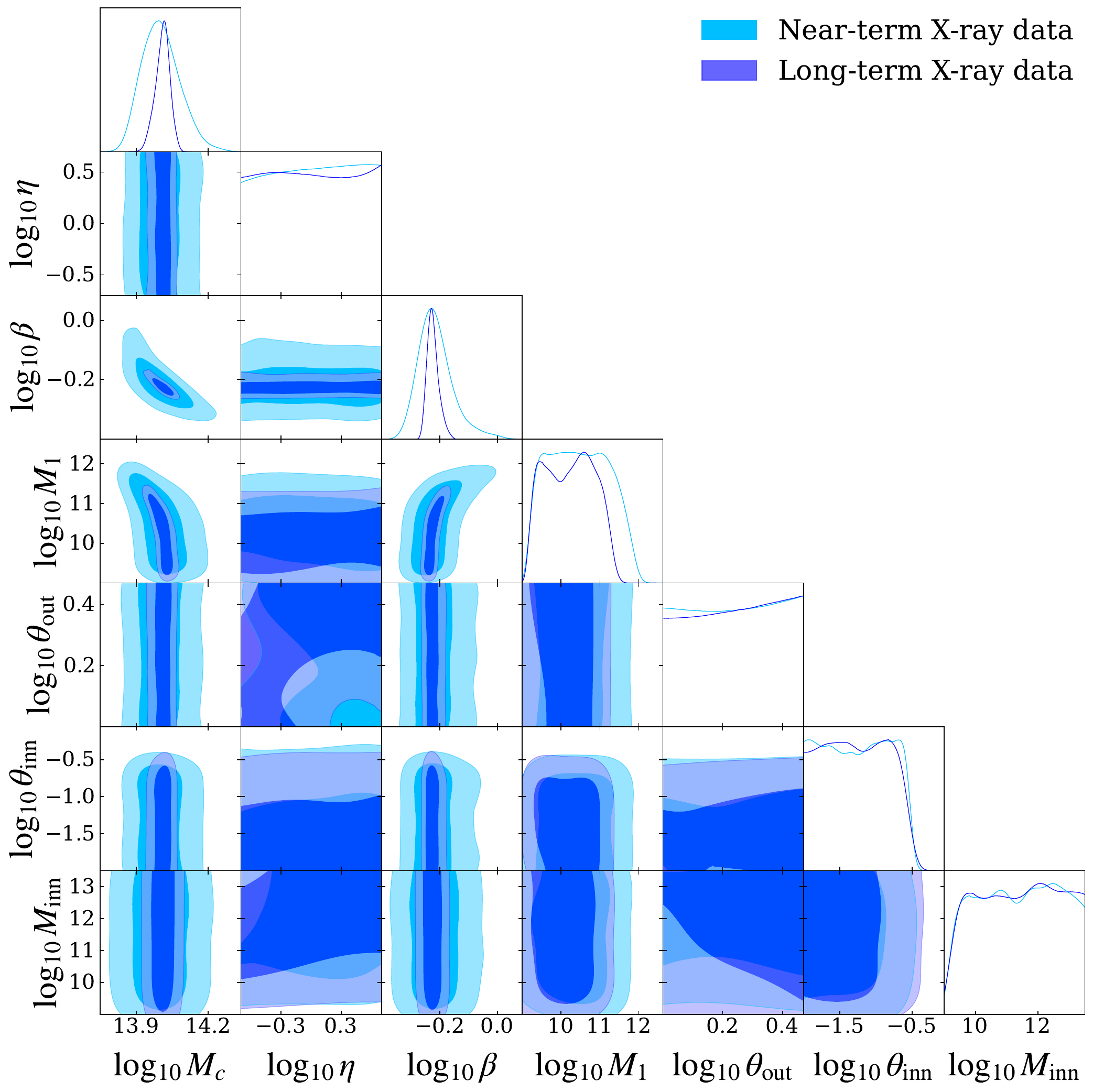}
      \caption{The marginalised posteriors on baryonic parameters obtained from mock measurements of the X-ray gas fraction for near-term data (light blue) and long-term data (dark blue), with the cosmological parameters kept fixed at the fiducial values used to generate the LSST-like cosmic shear data. We base the mock measurements for near-term data on 540 clusters from a combination of currently available SPT, HSC-XXL, and eROSITA data. The mock measurements for long-term data are derived assuming a sample of 5259 clusters from the full eROSITA dataset.}
      \label{fig:constraints_xrays}
    \end{figure}

\begin{figure}
      \centering
      \includegraphics[width=\linewidth]{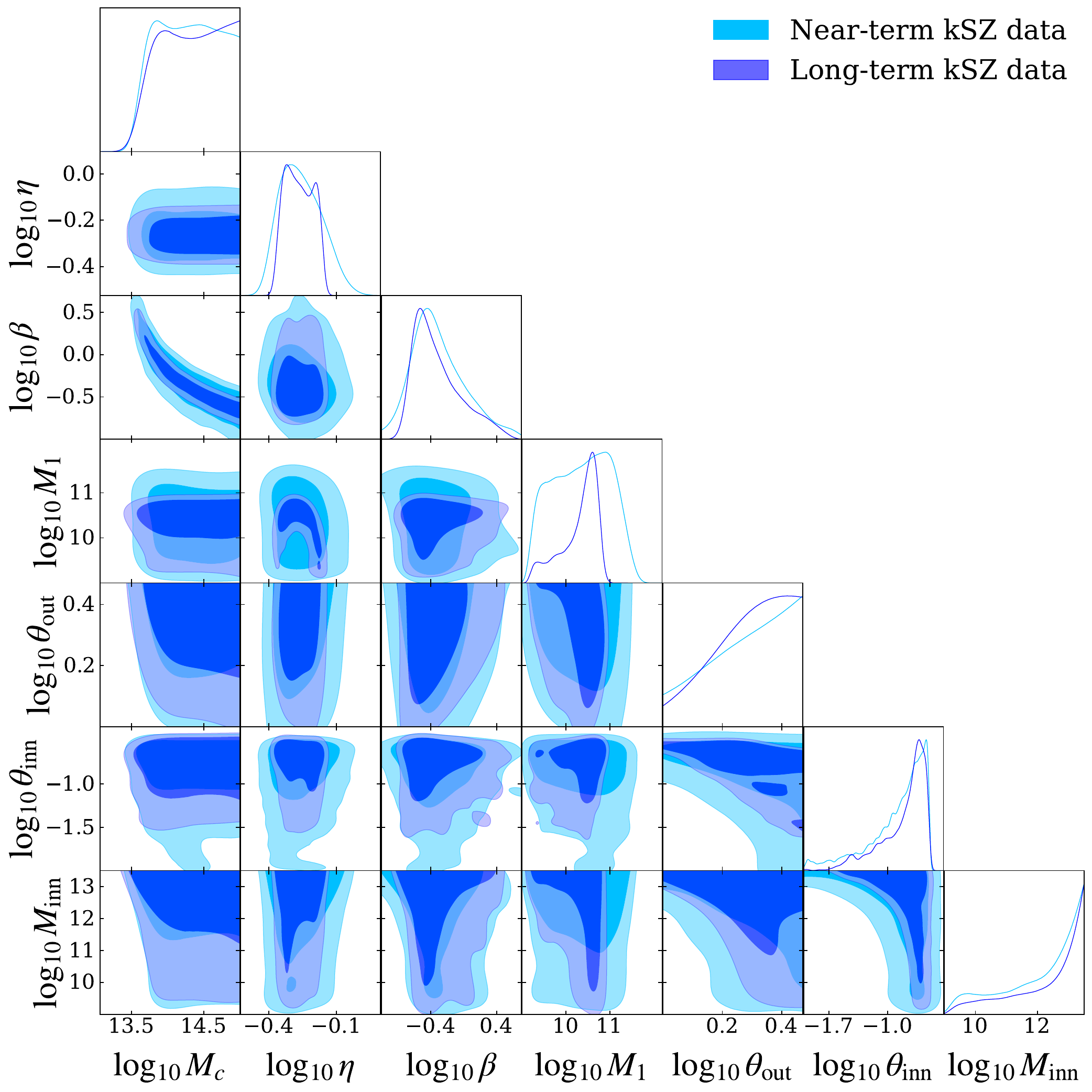}
      \caption{The marginalised posteriors on baryonic parameters obtained from mock measurements of the stacked kSZ profile for near-term data (light blue) and long-term data (dark blue), with the cosmological parameters kept fixed at the fiducial values used to generate the LSST-like cosmic shear data. We base the mock measurements for near-term and long-term data on the baseline noise level of the SO LAT and the CMB-S4 experiment, respectively.}
      \label{fig:constraints_ksz}
    \end{figure}

    With these mock datasets at hand, we run MCMC chains assuming fixed cosmological parameters and without weak lensing data. We do so for each dataset (X-ray or kSZ) separately, in order to explore their individual constraining power. The results are shown in Fig. \ref{fig:constraints_xrays} and Fig. \ref{fig:constraints_ksz} for the X-ray and kSZ datasets, respectively. The light blue and dark blue contours represent the marginalised posteriors obtained from the near-term and long-term datasets, respectively.

    We find that X-ray data primarily constrain $\log_{10} M_{\mathrm{c}}$ and $\log_{10} \beta$, whereas $\log_{10} \eta$ remains prior-dominated. In contrast to the X-ray signal, which traces only the bound gas, the kSZ signal is sensitive to the total gas profile. This broader sensitivity allows for constraints on $\log_{10} \eta$. However, the constraints on $\log_{10}M_c$ and $\log_{10}\beta$ from kSZ data are highly degenerate, and neither parameter can be constrained to sufficient precision. This is because the kSZ mock dataset assumed here probes only a single halo mass scale and redshift, which leads to a degeneracy between both parameters, since both control the abundance of bound gas. Hence, this prevents them from being individually constrained. This degeneracy is evident in the elongation of the posterior contours in the $\log_{10}M_{\mathrm{c}}-\log_{10}\beta$ panel in Fig. \ref{fig:constraints_ksz}. 

    The $68\%$ constraints on $\log_{10} M_{\mathrm{c}}$, $\log_{10} \beta$, and $\log_{10} \eta$ are listed in Table \ref{tab:bar_errors}. We include $\log_{10} \beta$ to highlight the sensitivity of the X-ray data to this parameter, despite its negligible effect on the degradation of cosmological constraints. We find that the posterior uncertainty on $\log_{10} M_{\mathrm{c}}$ improves by a factor of $\sim 2.5$ between the near-term and long-term X-ray mock datasets. The constraint on $\log_{10} M_{\mathrm{c}}$ derived from the long-term dataset is sufficient to achieve the calibration requirement found in the previous section of $\sigma(\log_{10} M_{\mathrm{c}}) \lesssim 0.1$. In turn, the constraints on $\log_{10} \eta$ improve by a factor $\sim 1.6$ from the near-term to the long-term kSZ samples. We find that the near-term mock dataset approaches the calibration requirement on $\log_{10} \eta$, while the long-term mock dataset satisfies this threshold. Hence, a joint analysis of WL with the long-term kSZ dataset ensures that cosmological constraints are not significantly degraded by marginalisation over $\eta$, whereas combining WL with the long-term X-ray dataset prevents degradation arising from uncertainties in $M_{\rm c}$. As mentioned above, these kSZ constraints could be significantly improved by assuming a more ambitious galaxy sample, covering a range of redshifts and halo masses. Note, however, that uncertainties in velocity reconstruction degrade our ability to use the total amplitude of the stacked kSZ measurements, which would affect the derived constraints on baryonic parameters \cite{Schaan:2021}. We leave a more detailed study for future work.

    Interestingly, none of the datasets are able to obtain sufficiently precise constraints on the other baryonic parameters, particularly $\log_{10} \theta_{\mathrm{inn}}$ or $\log_{10} M_1$, which, as we saw, can degrade the cosmological constraints on $h$ and $n_{\mathrm{s}}$. This is not surprising in the case of $\log_{10} M_1$, since neither probe is directly sensitive to the stellar component. However, it is interesting that, although kSZ measurements are sensitive to the scale dependence of the gas profile, they are not sufficient to constrain $\log_{10} \theta_{\mathrm{inn}}$ with sufficient precision. Other external probes, such as density profiles derived from X-ray data \citep{Grandis:2023}, or the cross-correlation between X-ray maps and other large-scale structure tracers \citep{2309.11129,2412.12081} could help constrain this parameter. Interestingly, we find that long-term kSZ data offers an improvement in the constraint on the stellar parameter, $\log_{10} M_1$. This sensitivity likely arises from the dependence of the baryon fraction of a given halo on the stellar mass fraction. An alternative explanation could involve the scale dependence of the kSZ signal at small radii, where stellar contributions become non-negligible. However, we exclude this possibility based on an analysis of the ratio of the kSZ profile as $M_1$ is varied relative to keeping $M_1$ fixed at its fiducial value. We find that the ratio remains effectively constant across all aperture radii examined, indicating that the constraining power is not driven by scale-dependent changes in the kSZ profile.

    \begin{table}
      \centering
      \begin{tabular}{cccc} 
        \hline
        Tracer & $\sigma(\log_{10} M_{\mathrm{c}})$ & $\sigma(\log_{10} \beta)$ & $\sigma(\log_{10} \eta)$\\
        \hline
        Near-term X-ray & 0.08 & 0.06 & prior-dominated \\
        Long-term X-ray & 0.03 & 0.02 & prior-dominated \\
        \hline
        Near-term kSZ & prior-dominated & prior-dominated & 0.08 \\
        Long-term kSZ & prior-dominated & prior-dominated & 0.05 \\
        \hline
      \end{tabular}
      \caption{The level to which we can constrain the baryonic parameters $\log_{10} M_{\mathrm{c}}$, $\log_{10} \beta$, and $\log_{10} \eta$ using X-ray gas fractions and stacked kSZ profiles as external tracers of the large-scale structure.}
      \label{tab:bar_errors}
    \end{table}

    \begin{figure}
      \centering
      \includegraphics[width=\linewidth]{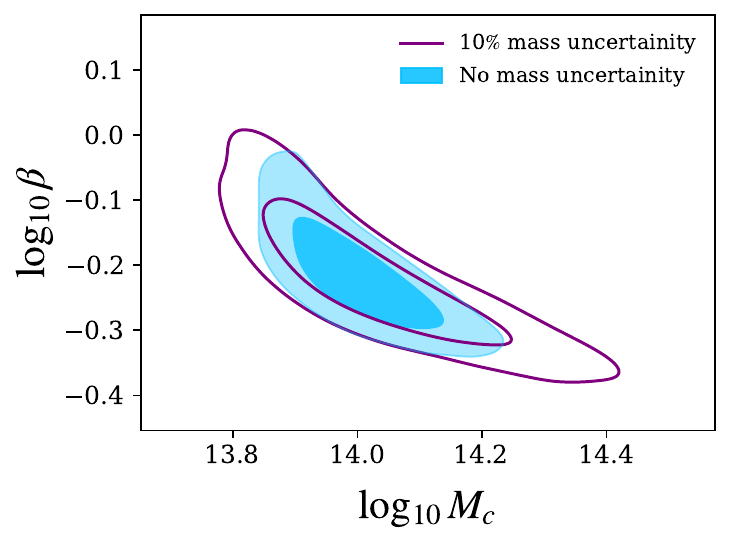}
      \caption{The constraints on $\log_{10} M_{\mathrm{c}}$ and $\log_{10} \beta$ from mock measurements of the X-ray gas fraction, based on near-term data for 540 clusters from a combination of SPT, HSC-XXL, and eROSITA. Blue contours represent constraints in the optimistic case without marginalisation over hydrostatic bias, while purple contours include marginalisation over a hydrostatic bias parameter with mass uncertainties of $\sim10\%$.}
      \label{fig:xray_hydrostatic_bias}
    \end{figure}

    To summarise, based on these results (summarised in Table \ref{tab:bar_errors}), and the requirements presented in the previous sections, we can draw the following conclusions:
    \begin{itemize}
      \item[(a)] Near-term X-ray data, based on the currently available data reported in \citet{Grandis:2023}, has the potential to constrain $\log_{10} M_{\mathrm{c}}$ to the level required for calibration, with long-term data further tightening the constraints. Additionally, $\log_{10} \beta$ is already tightly constrained with near-term data, contributing to reduced uncertainties in the derived cosmological parameters.
      \item[(b)] Near-term kSZ data, based on the baseline noise level of the SO LAT, is expected to meet the calibration requirement for $\log_{10} \eta$. Long-term CMB-S4 data will offer additional constraining power on $\log_{10} \eta$, further enhancing the overall cosmological parameter constraints. In both cases, these forecast constraints may be improved through the use of more ambitious galaxy samples, as well as refinements to velocity reconstruction.
      \item[(c)] Neither dataset is able to calibrate $\log_{10}M_1$ and $\log_{10}\theta_{\rm inn}$. Other external probes may be needed to break their degeneracy with cosmological parameters (in particular $h$ and $n_{\mathrm{s}}$). For instance, measurements of the thermal Sunyaev-Zel'dovich (tSZ) effect have the capability to constrain $\theta_{\mathrm{inn}}$ \citep[e.g.][]{Arico:2024}.
    \end{itemize}

    As a further investigation, we explore the effects of improvements to the galaxy catalogue. We combine the weak lensing dataset with the near-term X-ray and kSZ datasets, with the galaxy number density, $n_{\rm gal}$, in the near-term kSZ dataset increased by a factor of ten. Our results, displayed in the final row of Table \ref{tab:cosmo_errors}, show that the error on $S_8$ decreases by a factor of 1.2 relative to the WL + near-term X-ray + near-term kSZ dataset. Similar improvements are observed for $h$ and $n_{\rm s}$, with their respective errors decreasing by factors of 1.6 and 1.2.

    It is important to highlight that our results are optimistic in the sense that we have not accounted for systematic uncertainties in the modelling of the external tracers. In particular, a more realistic model for the stacked kSZ profile might discard compensated aperture photometry scales below $\sim1$ arcmin. Such a scale cut may be necessary to avoid small-scale modelling uncertainties (e.g. the impact of satellite galaxies). Moreover, we have interpreted the kSZ data as representing a sample of central galaxies at a fixed halo mass, consistent with the approach taken in recent kSZ analyses \citep{Bigwood:2024}. To model this, we introduce $\log_{10} M_{\rm halo}$ as a free parameter with a Gaussian prior centred at the value used to generate our previous stacked kSZ measurements, $\log_{10}(M_{\rm halo}/M_{\odot}) = \log_{10}(3 \times 10^{13})$. We investigate the extent to which these limitations affect the constraining power of the kSZ mock measurements for the near-term dataset. Our results indicate that these systematics do not significantly degrade the constraints on $\eta$, with similar conclusions holding for the long-term dataset. In particular, we find that the application of a scale cut and, removing scales $\theta_{\rm d}<1\,{\rm arcmin}$, degrades the constraints on $\log_{10}\eta$ by $\sim5\%$. Likewise, marginalising over the sample halo mass, assuming a $10\%$ prior uncertainty, leads to a similar increase in the posterior uncertainties on $\log_{10}\eta$. The combined impact of these systematics leads to a reduction in constraining power by a factor of $\sim1.1$. It is also important to note that stacked kSZ measurements are sensitive only to baryonic effects on halo masses corresponding to the host haloes of the galaxies that have been used in their measurements \citep{Lucie-Smith:2025}. Consequently, incorporating data from other baryonic probes, such as X-rays and the tSZ effect, is valuable for a more comprehensive analysis.
    
    On the X-ray side, our analysis does not include possible selection and mass measurement biases affecting X-ray clusters. We investigate this by introducing a free parameter $f$ to represent the discrepancy between the true and observed cluster masses, with $M_i = f M_i^{\rm true}$. This is effectively equivalent to accounting for a potential hydrostatic mass bias of the form $f = 1 - b_{H}$ \citep{1708.00697}. We then include the logarithm of the measured masses of each of the cluster samples used as additional elements of the data vector, with Gaussian calibration uncertainties of $\sigma(\log_{10}(M_i/M_\odot))=0.05$ (corresponding to a $\sim10\%$ uncertainty in the mass measurements). Our results, presented in Fig. \ref{fig:xray_hydrostatic_bias}, demonstrate that incorporating a $\sim 10\%$ mass measurement uncertainty degrades the constraining power of the near-term X-ray dataset on $\log_{10}M_{\mathrm{c}}$ by a factor of $\sim1.5$. Thus a tight control over selection effects and mass inference uncertainties is essential to obtain sufficiently precise constraints over baryonic parameters. Nevertheless, these results suggest that both kSZ and X-ray measurements remain powerful as complementary probes, even when these systematics are considered.

  \subsection{Cosmological constraints from cosmic shear with self-calibrated baryons}\label{ssec:results.cosmo}
    \begin{figure*}
      \centering
      \includegraphics[width=\linewidth]{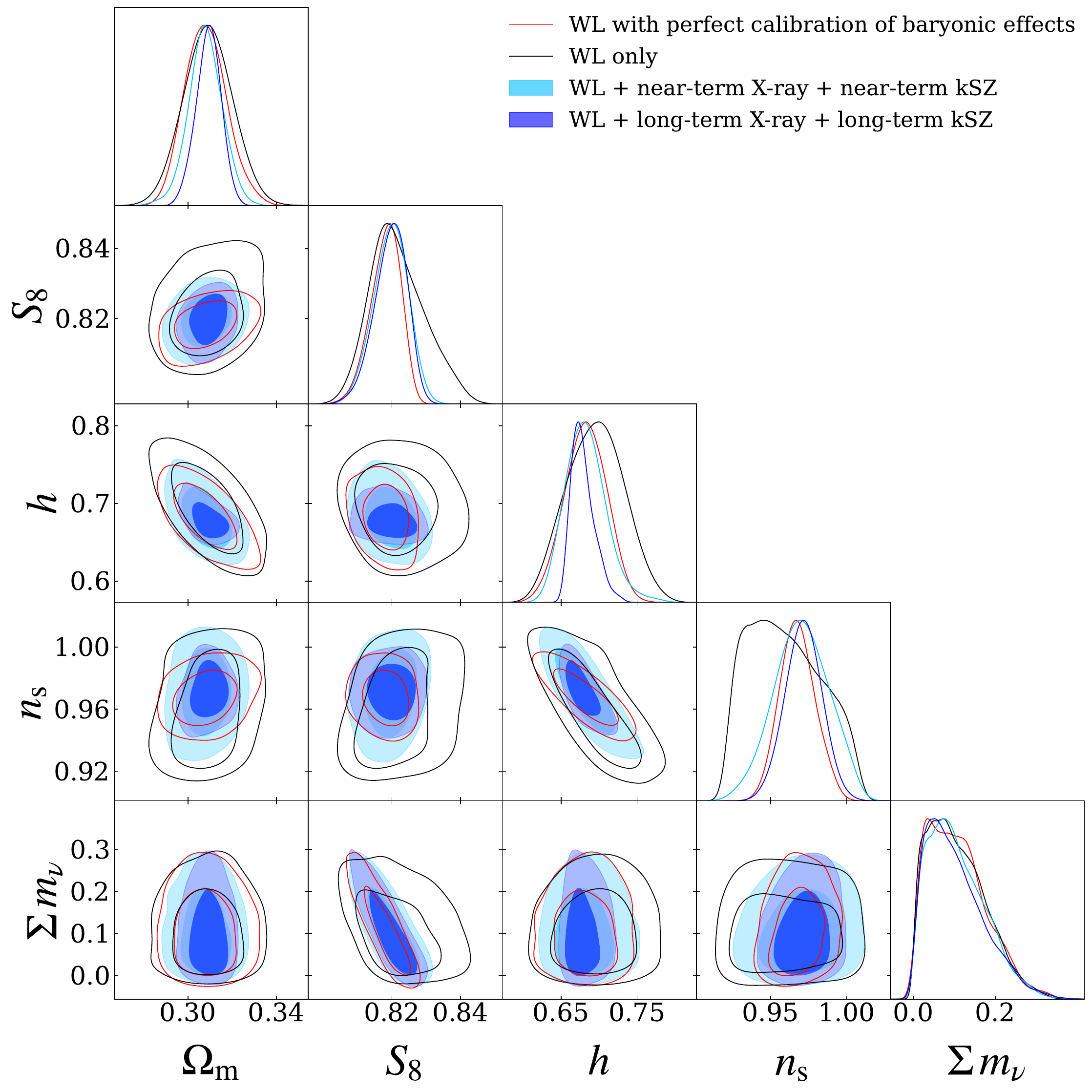}
      \caption{The marginalised posteriors for cosmological parameters for LSST-like weak lensing data only for fixed baryonic parameters (red), LSST-like weak lensing data only with marginalisation over baryonic parameters (black), and a joint analysis of LSST-like weak lensing data with near-term (light blue) and long-term (dark blue) external data. The inner and outer contours show the $68\%$ and $95\%$ confidence levels, respectively. We marginalise over intrinsic alignments, photometric redshift uncertainties, and multiplicative shape biases for the weak lensing data. The near-term X-ray and kSZ mock data is based on measurements from SPT, HSC-XXL, and eROSITA and the SO baseline noise curve, respectively. The long-term X-ray and kSZ mock data is based on future forecasts of eROSITA cluster samples and the S4-CMB experiment noise curve, respectively.}
      \label{fig:baryons_with_tracers}
    \end{figure*}

    \begin{figure}
      \centering
      \includegraphics[width=\linewidth]{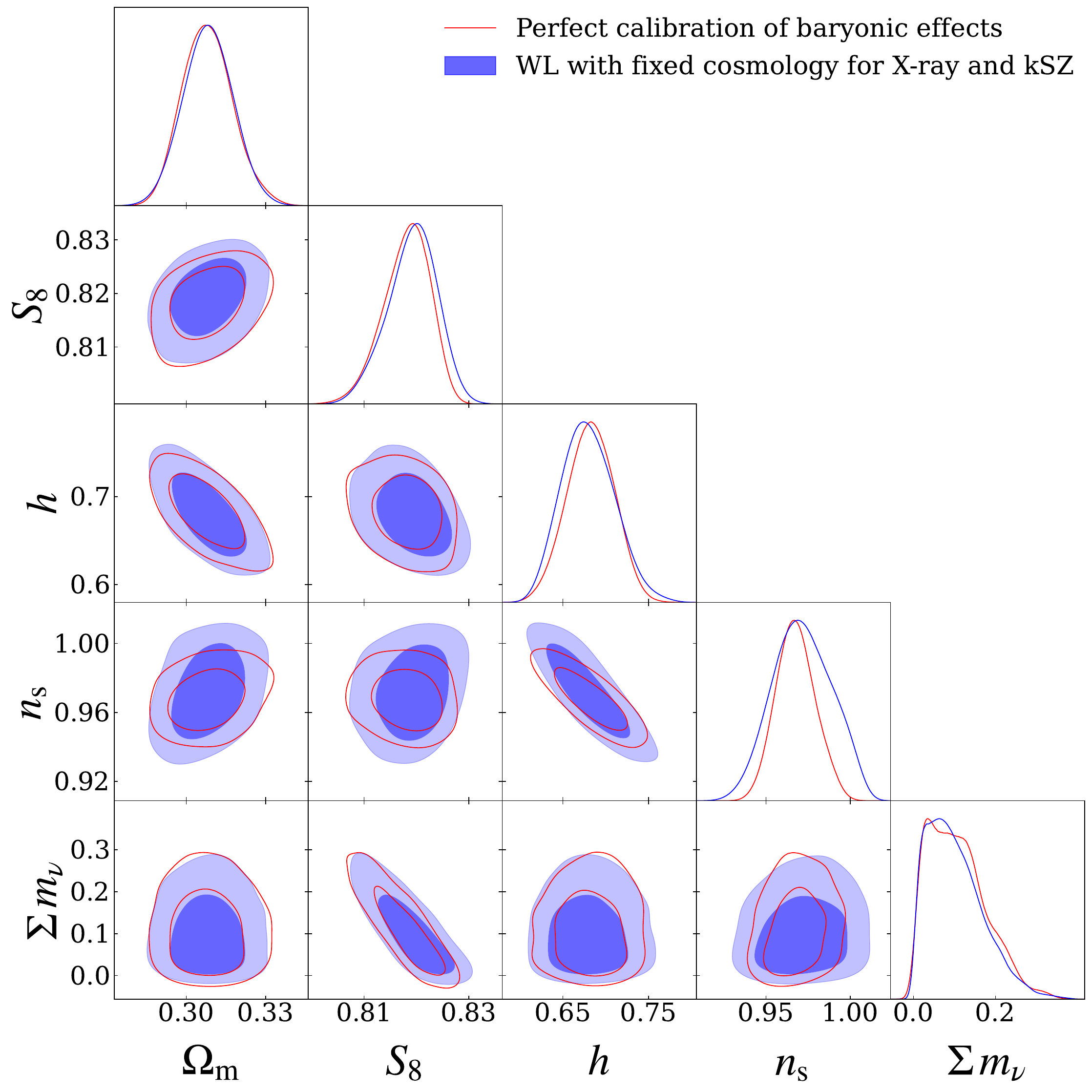}
      \caption{The marginalised posteriors for cosmological parameters derived from LSST-like weak lensing data. Results are shown for two cases: using weak lensing data alone with fixed baryonic parameters (red), and from a joint analysis combining weak lensing with long-term external datasets (i.e. X-ray and kSZ). In the external datasets, the cosmological parameters are fixed to the fiducial values used to generate the mock weak lensing data.}
      \label{fig:wl_fixed-cosmo-tracers}
    \end{figure}
    Having explored the ability of X-ray gas fractions and stacked kSZ measurements to calibrate baryonic parameters, we now derive the constraints that an LSST-like cosmic shear dataset could obtain on cosmological parameters when analysed in combination with these two external tracers in order to calibrate baryonic effects in a self-consistent likelihood.

    In Fig. \ref{fig:baryons_with_tracers}, we present an updated version of Fig. \ref{fig:info_loss_baryons}, now including the marginalised posteriors obtained from the combined WL + X-ray + kSZ data vector. The light blue and dark blue contours in Fig. \ref{fig:baryons_with_tracers} show the improvement in the recovery of cosmological constraints when self-calibrating baryonic effects by including external data, using near-term and long-term measurements, respectively. We find that the inclusion of long-term external datasets significantly tightens the constraints on all cosmological parameters when marginalising over baryonic effects (although, as noted in Section \ref{ssec:results.info_loss}, the neutrino mass sum constraints from cosmic shear alone remain unaffected by baryonic effects).
 
    In Table \ref{tab:cosmo_errors}, we present the marginal mean values and the corresponding $95\%$ confidence levels for the cosmological parameters $S_8$, $\Omega_{\mathrm{m}}$, $h$, and $n_{\mathrm{s}}$ evaluated across a range of tracer combinations. Our analysis demonstrates that incorporating LSST-like weak lensing data with X-ray and kSZ observations significantly mitigates the impact of baryonic effects on cosmological constraints. Specifically, the inclusion of near-term X-ray and kSZ data reduces the degradation factor from 1.9 to 1.2 for $S_8$, from 1.9 to 1.6 for $n_{\mathrm{s}}$, and from 1.4 to 1.0 for $h$. Furthermore, the uncertainty on $\Omega_{\rm m}$ is reduced by $\sim20\%$. Hence, we find that near-term data is already sufficient to restore the constraint on $h$ and $\Omega_{\rm m}$ to the level achieved under the assumption of perfect knowledge of baryonic feedback. Additional gains are achieved when forecasts for long-term X-ray and kSZ measurements are incorporated. This further reduces the degradation factor to 1.1 for $S_8$ and $n_{\mathrm{s}}$, while the constraints on both $h$ and $\Omega_{\rm m}$ are enhanced by a factor of 1.7 and 1.6, respectively with respect to the case of perfect calibration of baryonic effects with WL data alone. Long-term external calibrators thus have the potential to fully restore the cosmological constraining power of cosmic shear data in the presence of baryonic effects. The residual degradation in $n_{\rm s}$ and $S_8$ is likely due to the impact of the secondary BCM parameters $\theta_{\rm inn}$ and $M_1$, controlling the baryonic suppression factor on the smallest scales, as we noted in Section \ref{ssec:results.calreq}. As discussed in Section \ref{ssec:results.xrayksz}, the kSZ and X-ray observations considered here are not able to constrain these parameters, and other baryonic probes, such as tSZ and X-ray cross-correlations \citep{2412.12081}, or X-ray electron profile measurements \citep{Grandis:2023}, may be used to address this shortcoming. Moreover, other cosmological probes (e.g. CMB anisotropies) can place significantly tighter constraints on $n_{\rm s}$. These results highlight the value of integrating complementary probes of the large-scale structure in enhancing the precision of cosmological measurements.

    The improvement we observe in the constraints on $\Omega_{\mathrm{m}}$ and $h$ arises from the sensitivity of the kSZ measurements to cosmology via the baryon fraction controlling its normalisation, and the distance-redshift relation through the projected shape of the gas density profile. This additional constraining power may only be possible within the specific BCM parametrisation used here, and thus such constraints on $\Omega_{\mathrm{m}}$ and $h$ may not be achievable when employing a more general or less restrictive model. To verify that our main result, regarding the ability of external calibrators to restore the constraining power of cosmic shear in the presence of baryonic effects, is not affected by this, we repeat our analysis fixing the cosmological parameters used to predict the X-ray and kSZ measurements, while keeping them free for the cosmic shear power spectra. In this case, the sensitivity to cosmology of these tracers is cancelled, and they serve only to self-calibrate the baryonic parameters. Our results, presented in Fig. \ref{fig:wl_fixed-cosmo-tracers}, show that the improvement in the uncertainties on $\Omega_{\rm m}$ and $h$ for the long-term dataset in Fig. \ref{fig:baryons_with_tracers} disappears, and the resulting constraints converge to those obtained with cosmic shear data assuming a perfect knowledge of baryonic effects. Meanwhile, the residual degradation in the case of $S_8$ and $n_s$ observed earlier, remains in place (and in fact grows in the case of $n_s$).

    Finally, we note that, while combining weak lensing data with a single external tracer also enhances parameter estimates, the improvements are less substantial than those obtained from a joint analysis involving multiple tracers. Thus, it will be vital to combine complementary information from different probes of gas to fully self-calibrate the impact of baryons on weak lensing observables.


\section{Conclusions} \label{sec:conclusion}
  Baryonic feedback is the rearrangement of the matter density within and around dark matter haloes, driven by complex hydrodynamic processes. This phenomenon is one of the most significant sources of uncertainty in current weak lensing analyses, limiting the use of small-scale data for cosmological inference and potentially contributing to the tension between the $S_8$ measurements made by some weak lensing experiments and CMB data. Incorporating external tracers of the gas distribution in and around haloes provides a data-driven means to constrain the parameters that describe baryonic feedback and mitigate its impact on cosmological constraints.

  In this work, we present forecasts quantifying the precision required on baryonic parameters to recover the cosmological information that would be obtained from weak lensing data under the assumption of perfect knowledge of baryonic effects. Baryonic feedback is modelled using the \bemu emulator, which implements the baryon correction model, describing the gas content around haloes after being redistributed via feedback processes. We generate LSST-like weak lensing data up to multipoles $\ell_{\mathrm{max}} = 2000$ and examine the impact of marginalising over baryonic parameters to account for uncertainties arising from an incomplete understanding of feedback processes.

  Our analysis identifies $\log_{10} M_{\mathrm{c}}$ and $\log_{10} \eta$ as the baryonic parameters most responsible for the degradation of the cosmological constraints. The parameter $\log_{10} M_{\mathrm{c}}$ describe the abundance of bound, virialised gas, while $\log_{10} \eta$ characterises the ejected gas profile. We also find that the parameters characterising the fractional contribution from the stellar component, as well as the small-scale shape of the gas density profile, can also have a significant effect on the final constraints for future Stage-IV WL data. By marginalising over each baryonic parameter individually within a likelihood framework, we find that the baryonic parameters must be constrained at the level of $\sigma(\log_{10} M_{\mathrm{c}}) \sim \sigma(\log_{10} \eta) \lesssim 0.1$ to prevent degradation of cosmological constraints due to uncertainties associated with baryonic feedback. 

  Subsequently, we explore the potential of external tracers of the gas distribution to achieve the required calibration precision. Specifically, we consider measurements of the stacked kSZ profile and the bound gas fraction from X-ray observations of galaxy clusters. We find that X-ray gas fractions based on near-term measurements already satisfy the calibration requirement on $\log_{10} M_{\mathrm{c}}$ for cosmological constraints to be recovered, with long-term data expected to further tighten these constraints. Similarly, mock kSZ measurements based on near-term data achieve the necessary precision in $\log_{10} \eta$, while forecasts for upcoming experiments indicate that long-term kSZ observations will further enhance the constraints.

  The joint analysis combining LSST-like weak lensing data with X-ray and kSZ mock observations yields improved cosmological constraints compared to weak lensing alone, driven by the self-consistent calibration of baryonic parameters. To illustrate this, the combination with near-term external data decreases the factor by which the error on $S_8$ increases from $1.9$ to $1.2$, with long-term external data further decreasing this factor to $1.1$ (with the constraints on other parameters similarly improved). However, these results cannot be achieved by any individual gas probe. This demonstrates the potential of multi-tracer analyses to recover cosmological information degraded by baryonic effects. Future work could extend this framework by including additional tracers of the large-scale structure, such as the thermal Sunyaev-Zel'dovich (tSZ) effect \citep{2109.04458,Arico:2024}, fast radio bursts \citep[FRBs,][]{1901.02418,2201.04142,Reischke:2023,2411.17682,2506.08932}, and X-ray cross-correlations \citep{2204.13105,2309.11129,2412.04559, 2412.12081}.
  
  The analysis presented here suffers from a number of shortcomings. Our description of the kSZ stacking observable is relatively simplistic, and has ignored a number of potential sources of modelling uncertainty (e.g. velocity reconstruction uncertainties, the contribution of satellites and correlated structures, as well as subdominant mass-velocity correlations), as described in Section \ref{sssec:model.ksz}. Nevertheless, we have shown that our results are largely insensitive to uncertainties in the kSZ measurements on small scales, or to moderate uncertainties in the halo mass of the galaxy sample used for kSZ stacking. We have also used a relatively limited kSZ dataset, which will be improved by current and future data, by probing a range of redshifts and stellar masses. We have also not propagated uncertainties in the X-ray gas fraction measurements in our fiducial analysis, caused by selection effects or simplifying assumptions such as hydrostatic equilibrium. As we show in Section \ref{ssec:results.xrayksz}, these may have a critical impact on the ability of X-ray data to deliver the precision on baryonic parameters predicted here. The model used to describe baryonic feedback could also be easily generalised. For instance, more general BCM-like parametrisations have been proposed in the literature. Moreover, we have assumed that the BCM parameters do not have any redshift dependence (although we demonstrate in Appendix \ref{app:z_dependence} that allowing for redshift dependence in the most relevant baryonic parameters, $M_{\mathrm{c}}$ and $\eta$, does not significantly degrade the cosmological constraints obtained here). We have also focused our analysis on the $\Lambda$CDM framework, and studying the impact of external gas probes on the constraints achievable on extended models (e.g. considering an evolving dark energy equation of state) could be of interest given recent results from the latest measurements of type-Ia supernovae and baryon acoustic oscillations \citep{2401.02945,2503.14738}. Finally, our work has focused on quantifying the amount of information loss (i.e. the degradation in precision) incurred through uncertainties on baryonic effects, while ignoring the bias in the inferred cosmological parameters (i.e. the degradation in accuracy) that mis-modelling this contribution could lead to. External gas probes could be used to detect inconsistencies in the physical model that would lead to these biases, although the potential to do so will depend on the sensitivity of these probes, as well as the impact of their own observational and modelling systematics \citep{Kovac:2025}. Further work will be needed to address these questions.


\section*{Acknowledgements}
  We would like to thank Raul Angulo and Boryana Hadzhiyska for useful comments on earlier versions of this manuscript. AW is supported by a Science and Technology Facilities Council studentship.  DA and MZ acknowledge support from the Beecroft Trust. We made extensive use of computational resources at the University of Oxford Department of Physics, funded by the John Fell Oxford University Press Research Fund.

\section*{Data Availability}
  The data underlying this article will be shared on reasonable request to the corresponding authors.


\bibliographystyle{mnras}
\bibliography{references}



\appendix

\section{Redshift dependence of the baryonic parameters} \label{app:z_dependence}
  As discussed in Section \ref{sec:model}, the vanilla \bemu model does not incorporate potential redshift-dependent variations in the baryonic parameters. To assess the impact of this limitation, we examine the effect of introducing redshift dependence in the two most influential baryonic parameters, $M_{\mathrm{c}}$ and $\eta$, on cosmological parameter constraints. Specifically, we adopt the following parametrisations to model their redshift evolution
  \begin{equation}
    \log_{10} \widetilde{M}_{\mathrm{c}}(a) = \log_{10} M_{\mathrm{c}} + (1-a) \log_{10} M_{\mathrm{c},z}
  \end{equation}
  and
  \begin{equation}
    \log_{10} \widetilde{\eta} \, (a) = \log_{10} \eta + (1-a) \log_{10} \eta_{z},
  \end{equation}
  where $M_{\mathrm{c}}$ and $\eta$ are the original free parameters of the \bemu model, and $M_{\mathrm{c},z}$ and $\eta_z$ are additional free parameters introduced to capture potential redshift dependence.

\begin{figure}
    \centering
    \includegraphics[width=\linewidth]{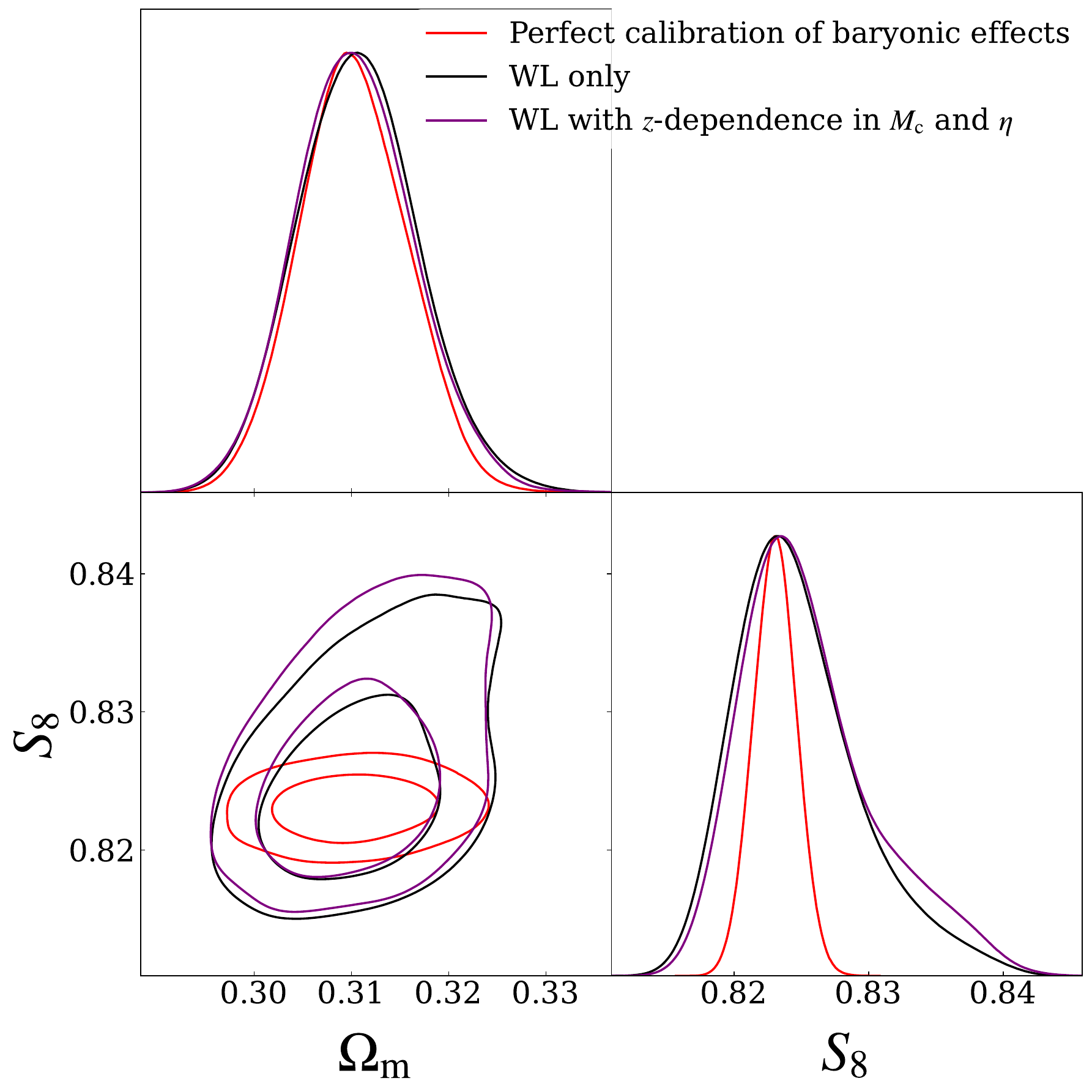}
    \caption{The marginalised posteriors of cosmological parameters derived from LSST-like weak lensing data up to multipoles $\ell = 2000$ for three cases. The red contours represent the ideal case, corresponding to perfectly calibrated baryonic effects. The black and purple contours correspond to constraints obtained by marginalising over baryonic parameters, without and with redshift dependence in $M_{\mathrm{c}}$ and $\eta$, respectively. In each case, the inner and outer contours denote the $95\%$ and $68\%$ confidence levels, respectively. All analyses consistently marginalise over intrinsic alignments, photometric redshift uncertainties, and multiplicative shape biases.}
    \label{fig:z_dependence}
\end{figure}

  Fig. \ref{fig:z_dependence} demonstrates the impact of incorporating redshift dependence in $M_{\mathrm{c}}$ and $\eta$ for a simplified two-parameter cosmological model. Our results indicate that introducing redshift evolution in in $M_{\mathrm{c}}$ and $\eta$ does not significantly degrade the cosmological constraints on $\Omega_{\rm m}$ and $S_8$ derived from weak lensing measurements.

\section{Implementation of \bemu} \label{app:bacco_implementation}

\subsection{Extrapolating in \textit{z}}

\begin{figure}
    \centering
    \includegraphics[width=\linewidth]{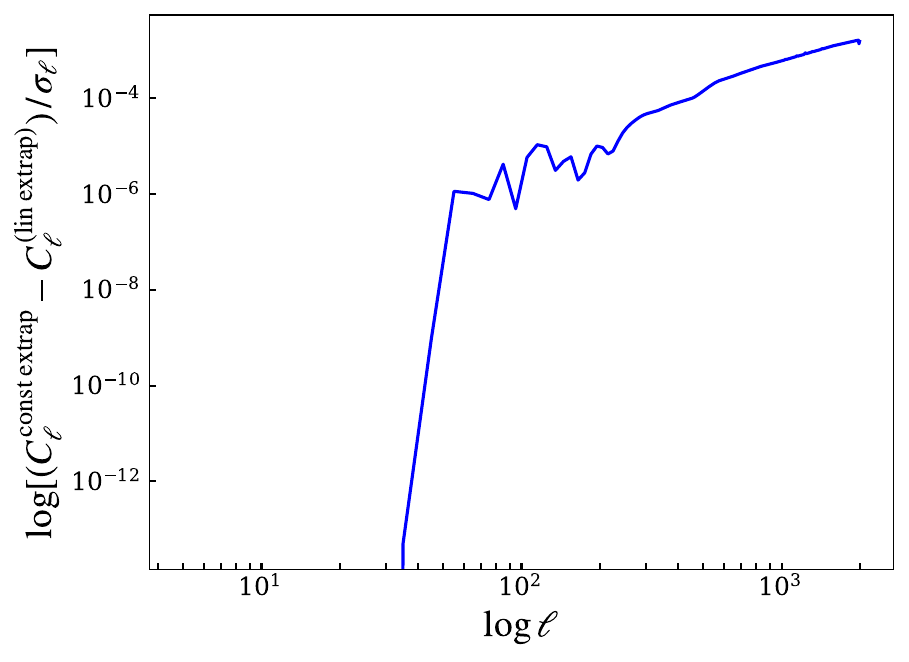}
    \caption{The difference between the angular power spectra for constant and linear extrapolation methods in the $z > 1.5$ regime for the auto-correlation of the fifth redshift bin. Here, $\sigma_\ell$ represents the square root of the diagonal of the covariance matrix for the angular power spectrum for the constant extrapolation approach.}
    \label{fig:beyond_bacco_z}
\end{figure}

Future cosmic shear surveys are anticipated to probe redshifts up to $z = 4.0$, however, \bemu is calibrated for redshifts up to $z = 1.5$. Consequently, it is necessary to extrapolate the non-linear matter power spectrum beyond the calibration range of \bemu in order to compute the angular power spectra for our LSST-like data at higher redshifts. To achieve this, we conducted \bemu simulations at redshifts $z=1.5$ and $z=2.0$ using the same cosmological and baryonic parameters as the mock data. This allows us to compare two extrapolation methods: (a) assuming a constant boost factor, $S_k(z=1.5)$, for $z > 1.5$, and (b) assuming a linear extrapolation of the boost factor for $z>1.5$. Our results are presented in Fig. \ref{fig:beyond_bacco_z}, which shows that the statistical uncertainty between the two extrapolation methods varies at most by $10^{-4} \sigma$. Hence, it is valid to assume constant extrapolation of the boost factor, $S_k(z=1.5)$, for $z > 1.5$. We find that this result is consistent across all redshift bin correlations. 

\subsection{Extrapolating in \textit{k}}

\begin{figure}
    \centering
    \includegraphics[width=\linewidth]{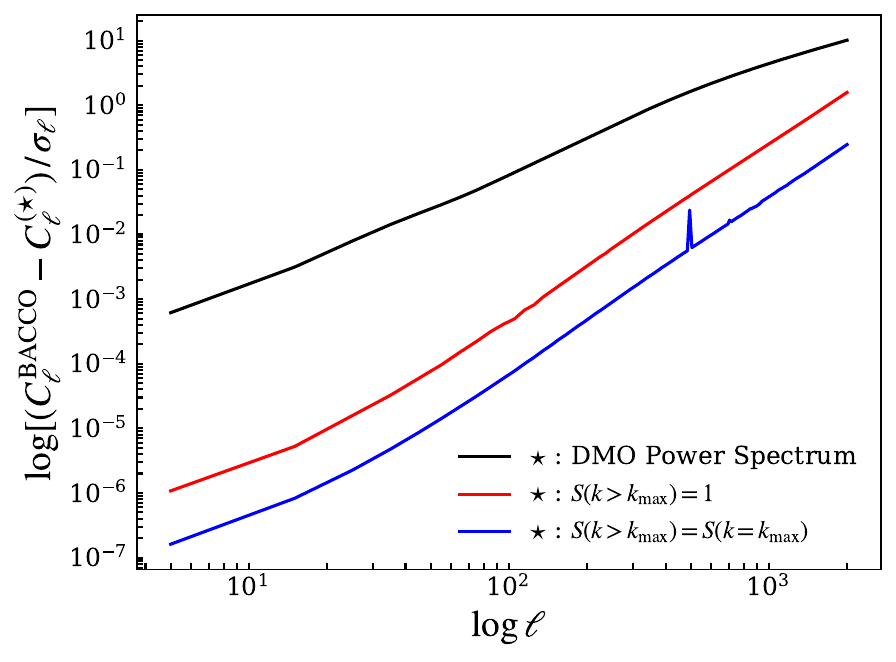}
    \caption{The angular power spectra for the different approximations in the regime $k > 3.0$, relative to the \bemu extrapolated angular power spectra for the cross-correlation between the first and third redshift bins. Here, $\sigma_\ell$ represents the square root of the diagonal of the covariance matrix for the \bemu extrapolated angular power spectrum.}
    \label{fig:beyond_bacco_k}
\end{figure}

Similarly, \bemu is calibrated up to wavenumbers of $k = 5 \, h \mathrm{Mpc}^{-1}$, which for our fiducial cosmology, corresponds to $k \sim 3 \, \mathrm{Mpc}^{-1}$. However, we use wavenumbers up to $k = 10 \, \mathrm{Mpc}^{-1}$ to generate our mock data. Hence, we assess the validity of the built-in extrapolation of the boost factor, $S(k)$, in \bemu for $k > 3 \, \mathrm{Mpc}^{-1}$ by comparing it to three different approximations in the $k > 3 \, \mathrm{Mpc}^{-1}$ regime. The approximations we considered were: (a) assuming the angular power spectrum follows the dark matter only spectrum for all $k$, (b) assuming the angular power spectrum follows the dark matter only spectrum for $k > 3 \, \mathrm{Mpc}^{-1}$ (i.e. $S(k > 3 \, \mathrm{Mpc}^{-1}) = 1$), (c) assuming the boost function remains constant at $S(k = 3 \, \mathrm{Mpc}^{-1})$ for $k > 3 \, \mathrm{Mpc}^{-1}$. Our results are presented in Fig. \ref{fig:beyond_bacco_k}. We observe that the difference between the built-in \bemu extrapolation in the regime and the dark matter only power spectrum in the regime $k > 3 \, \mathrm{Mpc}^{-1}$ is at most $0.1 \sigma$. Hence, it is valid to use \bemu in the regime $k > 3 \, \mathrm{Mpc}^{-1}$. We found that this conclusion holds consistently across all redshift bin correlations.

\bsp
\label{lastpage}
\end{document}